\documentclass[preprint]{aastex6}

\newcommand\erf{\rm erf}

\usepackage{mathtools}

\fullcollaborationName{The Friends of AASTeX Collaboration}

\begin{document}
\title{Effects of scattering, temperature gradients, and settling on the derived dust properties of observed protoplanetary disks}
\author{Anibal Sierra \altaffilmark{1} \& Susana Lizano \altaffilmark{1}}
\altaffiltext{1}{Instituto de Radioastronom\'ia y Astrof\'isica, UNAM, Apartado Postal 3-72, 58089 Morelia Michoac\'an, M\'exico}

\begin{abstract}
It is known that the millimeter dust thermal emission of protoplanetary disks is affected by scattering, such that for optically thick disks the emission decreases with respect to the pure absorption case  and the spectral indices can reach values below 2.  The latter can also be obtained with temperature gradients.
Using simple analytical models of  radiative transfer in thin slabs,  we quantify the effect of scattering, vertical temperature gradients, and dust settling on the emission and spectral indices of geometrically thin face-on accretion disks around young stars. We find that in vertically isothermal disks with large albedo ($\omega_{\nu} \gtrsim 0.6$), the emergent intensity can increase at optical depths between $10^{-2}$ and $10^{-1}$.  
We show that dust settling has important effects on the spectral indices in the optically thick regime, since the disk emission mainly traces small dust grains in the upper layers of the disk.  
The $\lambda = 870 \ \mu$m emission of these small grains can hide large grains at the disk mid plane when the dust surface density is larger than $\sim$ 3.21 g cm$^{-2}$. Finally, because of the change of the shape of the spectral energy distribution, optically thick disks  at 1.3 mm and grains with sizes between 300  $\mu$m $< a_{\rm max} <$ 1 mm have a  7 mm flux  $\sim 60$\% higher than the extrapolation from higher millimeter frequencies, assumed when scattering is neglected. This effect could provide an explanation to the excess emission at $\lambda = 7$ mm reported in several disks.
\end{abstract}


\section{Introduction}
Understanding the dust radiative transfer and dust opacity is fundamental to derive the dust properties in protoplanetary disks from the dust continuum observations. The role of the dust composition (e.g. \citealt{Pollack_1994}, \citealt{Yang_2019}), grain size and distribution (e.g. \citealt{Mathis_1977}; \citealt{Miyake_1993}; \citealt{Draine_2006}), polarization \citep{Kataoka_2015}, porosity and fluffiness (e.g. \citealt{Kataoka_2014}, \citealt{Tazaki_2019}) has been studied to determine the dust opacity properties. 
Since the dust opacity determines the dust temperature and the optical depth at different wavelengths,  knowing its properties allows to infer  the maximum grain size and the total dust mass from the observed disk emission.

In particular, the size of the dust grains is relevant in the theory of planet formation, where the dust grains are expected to collide, grow, and migrate during the first stages of the disk lifetime \citep{Brauer_2008}. Millimeter and centimeter dust grains have been inferred based on the spectral index of sub-millimeter and millimeter dust continuum observations (e.g. \citealt{Beckwith_1991}, \citealt{Jorgensen_2007}). In contrast, dust grains with sizes of only some hundred micrometers have been inferred from millimeter polarization observations \citep{Kataoka_2015}. It is not clear what is the origin of the disagreement between both methods, but some ideas have been proposed to address this problem.
For example, \cite{Yang_2019} found that both the polarization pattern and the small spectral index can be simultaneously explained by a population of dust grains with a maximum grain size of 3 mm and with a dust composition with pure absorptive carbonaceous. Their work shows that the typical dust compositions that include organics cannot reproduce small spectral indices and the polarization pattern simultaneously. However, if the organics are not included in the dust chemical composition, the small spectral index and the polarization pattern agree. This happens because the refractive index of the organics grains has an small imaginary part (the larger the imaginary part, the larger the absorption properties of the dust) compared with other grains (e.g. graphites, troilite, see Figure 4 of \cite{Yang_2019}). Then, the assumption of organics grains could be the responsible for the disagreement of the inferred maximum grain size between the spectral index and polarization method.  Thus, the assumption of a dust composition could play an important role when interpreting and inferring properties from observations.
Other authors have suggested (e.g. \citealt{Yang_2017}) that observations at the ALMA Band 7 ($\lambda \approx 870 \ \mu$m), where the polarization pattern is observed, are only tracing the small grains of hundreds of micrometers  in the upper layers of settled disks (i.e., that the emission at this band is optically thick). Then, observations at this wavelength cannot detect large grains in the mid plane, which can only be inferred at larger wavelengths, where the disk becomes optically thin.

By other hand, \cite{Miyake_1993} showed that, in the case of spherical particles with sizes similar to the observed wavelength, the scattering has important effects in the radiative transfer of the dust thermal emission. The parameter that describes the scattering regime is given by $x = 2\pi a/\lambda$, where $\lambda$ is the observed wavelength and $a$ is the size of the dust particles. The scattering process has two limits: the Rayleigh scattering regime $x<<1$, and the geometric scattering regime $x>>1$. Nevertheless, in protoplanetary disks, where the dust grains are expected to grow such to sizes of some millimeters or centimeters, $x \sim 1$  at sub-millimeter and millimeter  wavelengths. In this regime, opacities can be computed by the Mie theory for dust spherical grains, and it is found to be a major component of the total opacity. 
However, most of the modelling of the disk emission to obtain spectral indices has been done neglecting scattering effects in the solution of the radiative transfer equation (e.g. see review \citealt{Williams_2011}). Some recent examples of works where scattering was taken into account to study the dust disk properties are \cite{Soon_2017}, \cite{Carrasco_2019}, \cite{Liu_2019}, \cite{Sierra_2019}.

Nevertheless, dust growth in protoplanetary disks is believed to generate dust grains with non-spherical geometries. This has been proven to occur in laboratory experiments (e.g. \citealt{Blum_2000}) and numerical simulations (e.g. \citealt{Wada_2007}, \citealt{Wada_2008}). Such fluffly grains, which are naturally expected to form, were also proposed by \cite{Kataoka_2013} as a method to avoid the radial drift barrier.
If dust grains are fluffly, the Mie theory is no longer valid (see the methods developed in \cite{Tazaki_2016} for fractal dust aggregates), and dust scattering opacity can be neglected, since most of the photons are forward scattered \citep{Tazaki_2019}. However, observations of the polarization pattern due to dust self scattering mentioned above (e.g. \citealt{Kataoka_2016}, \citealt{Kataoka_2017}, \citealt{Stephens_2017}, \citealt{Ohashi_2018}, \citealt{Bacciotti _2018}), and the anomalous low spectral indices (e.g. \citealt{Dent_2019}, \citealt{Liu_2019}) suggest that scattered light has important effects on  the observed millimeter disk emission.  Therefore, dust grains are likely compact spheres instead of fluffly aggregates. The inference of such compact grains has constrained the dust grain composition and fluffiness, and there is now a consensus that the radiative transfer of the dust emission must consider the scattering effects when evaluating the emergent intensity from the disks.

One of the first works which compute the effects of the scattering on the radiative transfer equation was done by \cite{Miyake_1993}.
They found an analytic solution for the mean intensity $J_\nu$ of a plane-parallel slab including scattering, by using the Eddington approximation. This solution was used by \cite{Dalessio_2001}  to obtain the source function $S_\nu$ and integrate the radiative transfer equation in models of accretion disks around T Tauri stars.
Recently, the Miyake solution was included in radiative transfer models of the complex substructures observed in dust continuum images with high angular resolution of disks from the DSHARP project \citep{Birnstiel_2018}. \cite{Liu_2019} also used the Miyake solution together with Monte Carlo simulations to explain the anomalous spectral indices reported in some optically thick protoplanetary disks. He found that very low spectral index cannot be explained by pure absorption models. However, these low spectral indices can be produced by scattering effects in optically thick disks with high albedo.
Furthermore, \cite{Zhu_2019} tested the Miyake solution by comparing their results with Monte Carlo radiative transfer simulations. They found that the plane parallel Miyake approximation is in agreement with the simulations. In their work, the inferred optical depths in the DSHARP disks at $\lambda = $1.3 mm ($\tau_{\rm obs} \sim 0.6$) can be naturally explained if the emergent intensity (taking into account the scattering) is wrongly interpreted as a non-scattering process.
They also showed that the non-scattering interpretation of the ALMA observations could be hiding dust mass in the disks by one order of magnitude. In the case of the resolved HL Tau disk, \cite{Carrasco_2019} determined that the estimation of the disk mass varies from $\sim 0.5 \times 10^{-3} M_{\odot}$ in the pure absorption case, to $1.0 \times 10^{-3} M_{\odot}$ in the scattering case. Then, neglecting the scattering effects in the HL Tau disk means that the disk mass is underestimated by a factor of $\sim 50 \%$.
For unresolved observations of disks, one can only obtain a lower limit of the mass  assuming a dust temperature (e.g. \citealt{Ansdell_2016}).

Furthermore, based on the accretion histories of young stellar objects, \cite{Liu_2017} and \cite{Hartmann_2018} suggested that the disk masses have been underestimated by at least one order of magnitude. In addition, more massive disks would naturally have the solid mass reservoir required for the known exoplanets in Class II disks \citep{Najita_2014}.

In this paper we discuss the effects of scattering in the emergent intensity of geometrically thin protoplanetary disks, the spectral indices at radio frequencies, and the effects of temperature and settling. Section \ref{SEC:dust_opacity} summarises the dust opacity properties at different radio wavelengths for grains of different sizes; these results are then used in the following sections to describe their effects in the emergent intensity of thin slabs. 
In Section \ref{SEC:Radiative_transfer} we discuss the solution to the radiative transfer equation taking into account the scattering effects, which was already presented in \cite{Sierra_2019}, and it is used in this work to compute and compare the emergent intensity in the cases with and without scattering.

In Section \ref{SEC:spectral_index} we discuss how scattering modifies the spectral indices at different wavelengths with respect to the pure absorption case as a function of the optical depth and the maximum grain size. Slices to this plane in the optically thin and thick regime are found to coincide with the results from \cite{Zhu_2019}. In the above sections, we assumed a vertically isothermal and a non-settled disk. These assumptions are no longer taken into account in the following sections, where the effects of the vertical gradient temperature (Section \ref{SEC:Temperature}) above the mid plane and dust settling (Section \ref{SEC:settling}) also modify the spectral indices in both the scattering and pure absorption cases.

Section \ref{SEC:SED} compares the spectral energy distribution (SED) taking or not into account the scattering effects for different disk inclinations, and in Section \ref{SEC:7mm} we explain how the non-scattering assumption in the radiative transfer equation can lead to a wrong interpretation of an excess emission at $\lambda = 7$ mm reported in several disks. Conclusions are presented in Section \ref{SEC:conclusions}.

\section{Dust opacity}
\label{SEC:dust_opacity}
The main source of opacity in protoplanetary disks is dust. These particles, which are expected to have sizes from 0.05 $\mu$m to some millimeter or centimeter, are the responsible of absorbing and emitting radiation at almost all the disk spectrum.
The dust opacity has two contributions: absorption and scattering. The extinction coefficient $\chi_{\nu}(a)$ ($\rm{cm}^2/\rm{g}$ in cgs units) for grains with radius $a$ at the frequency $\nu$ is the sum of the absorption coefficient $\kappa_{\nu}(a)$, and the scattering coefficient $\sigma_{\nu}(a)$
\begin{equation}
\chi_{\nu}(a) = \kappa_{\nu}(a) + \sigma_{\nu}(a),
\label{EQ:Ext_a_def}
\end{equation}
while the albedo $\omega_{\nu}(a)$ is defined as the ratio between the scattering coefficient and the extinction coefficient
\begin{equation}
\omega_{\nu}(a) = \frac{\sigma_{\nu}(a)}{\chi_{\nu}(a)}.
\end{equation}

The scattering coefficient $\sigma_{\nu}(a)$ is these equations is the effective scattering coefficient, which is defined as
\begin{equation}
\sigma_{\nu}(a) = (1-g_{\nu}(a)) \sigma_{\nu}^{\rm single}(a),
\label{EQ:effective_scattering}
\end{equation}
where $\sigma_{\nu}^{\rm single}(a)$ is the single scattering coefficient and $g_{\nu}(a)$ is the asymmetry parameter defined as the expectation value of the cosine of the scattering angle. The correction $g_{\nu}(a)$ is included to take into account for the non-isotropic scattering \citep{Henyey_1941}. For example, if the probability that a photon can be scattered in the forward direction is larger than in any other direction, the value of $g_{\nu}(a)$ is closer to 1 because it is equivalent to a less effective scattering. Thus, $\sigma_{\nu}(a)$ decreases according to equation (\ref{EQ:effective_scattering}). For isotropic scattering, the asymmetry parameter is $g_{\nu}(a) = 0$.

The coefficients $\kappa_\nu(a)$, $\sigma_\nu^{\rm single}(a)$, and $g_\nu(a)$  depend on the grain size and the observed frequency and can be written in terms of the dust dielectric constants (eqs. 2.55-2.57 of  \citealt{Kruegel_2003}). In this work, the coefficients are computed using the Mie theory for spherical dust grains of a given composition. We adopt the \cite{Pollack_1994} dust abundances (26\% silicates, 31\% organics, 43\% ice) and use the \cite{Dalessio_2001} code to compute these coefficients as a function of the grain radius.

In the interstellar medium (ISM) the dust grains follow a particle size distribution \citep{Mathis_1977} $n(a)da \propto a^{-p}da$, which gives the number density of dust grains with radius between $a$ and $a+da$. The ISM is characterized by a slope $p=3.5$, which is also typically assumed in protoplanetary disks. However, if the coagulation process is dominant and dust growth occurs, a smaller value for the slope is expected \citep{Miyake_1993}.

The mass weighted monochromatic coefficients are integrated between the minimum grain size $a_{\rm min}$  and the maximum grain size $a_{\rm max}$, for example, for the absorption coefficient
\begin{equation}
\kappa_{\nu} = \frac{\int_{a_{\rm min}}^{a_{\rm max}} \kappa_{\nu}(a) a^3 n(a)da }{\int_{a_{\rm min}}^{a_{\rm max}}  a^{3}n(a)da},
\label{EQ:average_opacities}
\end{equation}

The coefficients $\kappa_{\nu}$, $\sigma_{\nu}$, $\chi_{\nu}$ are also called opacity coefficients, because they absorb and/or scatter the radiation from the line of sight. The albedo is then defined as $\omega_{\nu} = \sigma_{\nu}/\chi_{\nu}$.

The opacity coefficients and albedo as a function of the frequency are the same than those shown in top panels in Figure 4 of \cite{Carrasco_2019} for different maximum grain sizes. In this work, we focus in the dust properties at millimeter wavelengths, specifically between $\lambda = 0.87$ and 7 mm, where the absorption, scattering, and extinction coefficients can be fitted by a power law of the frequency as
\begin{eqnarray}
\kappa_{\nu} = \kappa_{0}\left(\frac{\nu}{\nu_0}\right)^{\beta_{\kappa}}, \quad \quad &
\sigma_{\nu} = \sigma_{0}\left(\frac{\nu}{\nu_0}\right)^{\beta_{\sigma}}, \quad \quad &
\chi_{\nu} = \chi_{0}\left(\frac{\nu}{\nu_0}\right)^{\beta_{\chi}},
\end{eqnarray}
where $\beta_{\kappa}$, $\beta_{\sigma}$, and  $\beta_{\chi}$ are the opacity spectral indices, and $\kappa_0$, $\sigma_0$,  and $\chi_0$ are the absorption, scattering, and extinction coefficients at the reference frequency $\nu_0$. 
Note that $\beta_{\chi}$ depends on the relative magnitudes of the absorption and scattering coefficients. $\beta_{\chi} \rightarrow \beta_{\sigma}$ if $\sigma_{\nu} >> \kappa_{\nu}$, and $\beta_{\chi} \rightarrow \beta_{\kappa}$ if $\kappa_{\nu} >> \sigma_{\nu}$. 
By definition, 
\begin{equation}
\omega_{\nu} = \omega_{0}\left(\frac{\nu}{\nu_0}\right)^{\beta_{\omega}}.
\end{equation}
where $\beta_{\omega} = \beta_{\sigma} - \beta_{\chi}$.

The opacity spectral indices are the same than those shown in bottom panels in Figure 4 of \cite{Carrasco_2019}, where the slope $p$ of the particle size distribution is set to $p=3.5$ and the minimum grain size is set to $a_{\rm min} = 0.05 \ \mu$m. The latter value does not affect the opacity properties when $a_{\rm max} >> a_{\rm min}$ \citep{Draine_2006}. Note that the albedo rapidly increases when $a_{\rm max} \sim 100 \ \mu$m, and it reaches a large ($\omega_{\nu} \lesssim 1$) and constant ($\beta_{\omega} \sim 0$) value in the range $200 \  \mu {\rm m}  \lesssim a_{\rm max} \lesssim 1$ cm. Therefore, the dust opacity is dominated by the scattering coefficient at mm wavelengths if $a_{\rm max}$ is in the order of mm or cm.

The opacity spectral indices are fundamental to interpret the dust continuum millimeter observations. A common practice is to neglect the scattering opacity in the solution of the emergent intensity $I_{\nu}$ in the radiative transport equation, such that $I_{\nu} = B_{\nu}(T) (1 - e^{-\tau_{\kappa_\nu}})$, where $B_\nu(T)$ is the Planck function and $\tau_{\kappa_\nu}$ is the optical depth associated to the absorption coefficient. In addition, if the emission is within the Rayleigh-Jeans regime ($B_{\nu}(T) \propto \nu^2$) and in the optically thin limit $(1 - e^{-\tau_{\kappa_\nu}} \approx \tau_{\kappa_\nu} \propto \nu^{\beta_{\kappa}})$, the emergent intensity can be written as $I_{\nu} \propto \nu^{2 + \beta_{\kappa}}$. Then, fitting a power law to the observed $I_{\nu}$, one obtains $\beta_{\kappa}$ to infer the size of the dust grains. Many authors have questioned the Rayleigh-Jeans and optically thin assumptions; nevertheless, the main problem is that the scattering effects are not negligible at mm wavelengths for mm-cm dust grains.

So, for a given observed value of $\beta_{\rm obs}$, the interpretation of the maximum grain size varies if it is compared with $\beta_{\chi}$ or $\beta_{\kappa}$. For example, for an observed value of $\beta_{\rm obs} = 0.5$, the value of $a_{\rm max}$ could vary around two orders of magnitude depending on which curve is used to interpret $\beta_{\rm obs}$. When scattering is not taken into account, the curve of $\beta_{\kappa}$ is used. When scattering effects are included, the curve of $\beta_{\chi}$ should be used taking also into account the modification of the source function from the typical Planck function.

Note also that $\beta_{\kappa}$ only varies between $\sim 1.3$ ($a_{\rm max} \sim 10$ cm) and $\sim 2.5$ ($a_{\rm max} \sim 400 \  \mu$m); however, $\beta_{\chi}$ varies from $\sim 0.5$ ($a_{\rm max} \sim $ mm, cm) to $\sim 4$ ($a_{\rm max} \sim 200 \ \mu$m).  Nevertheless, when the maximum grain size is very small ($a_{\rm max}<< 100 \ \mu$m), the albedo is close to $0$, and both $\beta_{\chi}$, $\beta_{\kappa}$ converge to 1.7, which is the typical value of the ISM \citep{Draine_2006}.

\section{Radiative Transfer}
\label{SEC:Radiative_transfer}
The change of the specific intensity while radiation travels through matter follows the radiative transfer equation
\begin{equation}
\frac{dI_{\nu}}{d\tau_{\chi_\nu}} = - I_{\nu} + S_{\nu}(T),
\label{EQ:RadTrans}
\end{equation}
where $S_{\nu}$ is the source function and $\tau_{\nu}$ is the optical depth along the line of sight, which can be computed from
\begin{equation}
\frac{d\tau_{\chi_\nu}}{dZ} = \chi_{\nu} \rho_{\rm d},
\label{EQ:OptDepth}
\end{equation}
and where $Z$ is the distance in the line of sight though the dust density $\rho_{\rm d}$. 
An analytical solution of the source function with scattering of a thin slab was obtained by \cite{Miyake_1993} for a thin slab. In their work, the emergent intensity is computed using the two-stream approximation. Their solution of the source function has been used by other authors to compute the emergent intensity  of protoplanetary disks   including scattering  (e.g. \citealt{Dalessio_2001}; \citealt{Birnstiel_2018}). In particular, given the Miyake scattering source function, an analytical integration of the transfer equation is possible in the case of a face-on vertically isothermal thin slab \citep{Sierra_2019}. In this and the next section, we use the latter solution for the emergent intensity, which in terms of the  true absorption optical depth $ \tau_{\kappa_\nu} $ and albedo $\omega_\nu$, it can be written as
\begin{equation}
I_{\nu}^{\mathrm{sca}} = B_{\nu}(T) \left[ 1 - \exp\left(-\frac{\tau_{\kappa_{\nu}}}{1-\omega_{\nu}}\right) + \omega_{\nu} {\cal F} (\tau_{\kappa_{\nu}},\omega_\nu) \right],
\label{EQ:Intensity_scattering}
\end{equation}
where
\begin{eqnarray}
{\cal F} (\tau_{\kappa_{\nu}},\omega_{\nu}) &=& \frac{1}{\left( \sqrt{1-\omega_{\nu}} - 1 \right) \exp\left(-\sqrt{\frac{3}{1-\omega_{\nu}}} \tau_{\kappa_{\nu}} \right)  - \left(\sqrt{1-\omega_{\nu}}+ 1 \right)}  \times  \nonumber\\
 & & \left\lbrace  \frac{1 - \exp\left[-(\sqrt{3(1-\omega_{\nu})}+1) \frac{\tau_{\kappa_{\nu}}}{1-\omega_{\nu}} \right] }{ \sqrt{3(1-\omega_{\nu})} +1}        +      \frac{\exp \left[ - \frac{\tau_{\kappa_{\nu}}}{1-\omega_{\nu}} \right] - \exp\left[ -\sqrt{\frac{3}{1-\omega_{\nu}}} \tau_{\kappa_{\nu}} \right]    }{\sqrt{3(1-\omega_{\nu})} - 1}    \right\rbrace,
\end{eqnarray}
and $I_{\nu}^{\rm sca}$ is the emergent intensity that takes into account both the absorption and the scattering effects. Note that $\tau_{\kappa_\nu}$ and $\omega_\nu$ are independent variables.
If scattering is neglected ($\omega_{\nu} = 0$), then equation (\ref{EQ:Intensity_scattering}) reduces to the well known pure absorption emergent intensity
\begin{equation}
I_{\nu}^{\mathrm{abs}} = B_{\nu}(T) \left[ 1 - \exp\left(- \tau_{\kappa_{\nu}} \right) \right].
\label{EQ:Intensity_absorption}
\end{equation}
In this work, we use a thin slab model to examine the expected SEDs and spectral indices of geometrically thin accretion disks around young stars. Here after, we will refer to it as a disk model.

We define the ratio between the scattering and absorption emergent intensities as
\begin{equation}
{\cal R}_{\nu} = \frac{I_{\nu}^{\mathrm{sca}}}{I_{\nu}^{\mathrm{abs}}}.
\label{EQ:R}
\end{equation}
Note that this is not the same definition given by equation (10) of \cite{Zhu_2019}. In their definition, the ratio is with respect to the black-body emission, which equals the emergent intensity without scattering only in the optically thick regime. However, in our definition, the ratio directly gives the increase or decrease of the emergent intensity when scattering is included in the radiative transfer equation. The ratio defined in \cite{Zhu_2019} reduces to unity for $\omega_{\nu} = 0$ only for optically thick disks (see Figure 1 of \cite{Zhu_2019}). Instead, ${\cal R}_\nu \rightarrow 1$ for $\omega_\nu =0$, in both the optically thick and thin regime.

From now on, when we mention the optical depth regime (as thin or thick), we will refer to absorption optical depth $\tau_{\kappa_{\nu}}$. Note that even if $\tau_{\kappa_{\nu}} << 1$, $\tau_{\chi_{\nu}}$ can be large if $\omega_{\nu} \sim 1$. The ratio ${\cal R}_{\nu}$ in the optically thin and optically thick limits is given by \cite{Sierra_2019}
\begin{equation}
{\cal R}_{\nu} =
\begin{cases*}
1 & , $\tau_{\kappa_{\nu}} <<1$ \\
1 - \frac{\omega_{\nu}}{(\sqrt{1 - \omega_{\nu}} + 1 ) (\sqrt{3(1-\omega_{\nu})} + 1)}       & , $\tau_{\kappa_{\nu}} >>1$. 
\end{cases*}
\label{EQ:Optical_thick_limit}
\end{equation}

\begin{figure}
\centering
\includegraphics[scale=1.2]{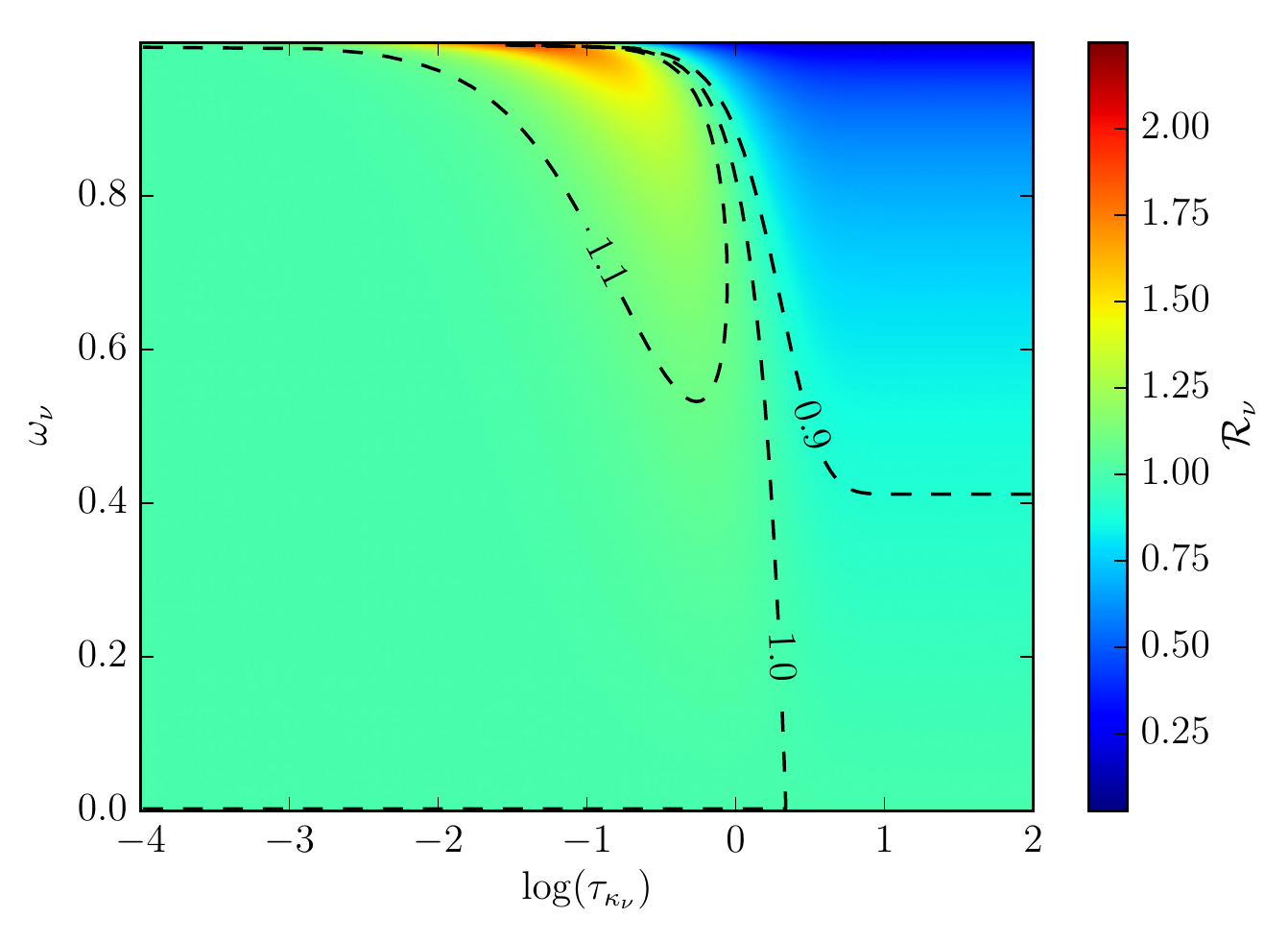}
\caption{Ratio between the emergent intensity with scattering effects and without scattering effects as a function of the optical depth associated to the absorption and the albedo. The isocontours show the region where $\cal{R}_{\nu}$ is 0.9, 1.0, and 1.1.}
\label{FIG:Ratio_Intensities}
\end{figure}

Figure (\ref{FIG:Ratio_Intensities}) shows the ratio ${\cal R}_{\nu}$ as a function of $\tau_{\kappa_{\nu}}$ and $\omega_{\nu}$. The isocontours where ${\cal R}_{\nu} = $ 0.9, 1, and 1.1 are plotted as dashed lines as reference. Note that the values $0.9<{\cal R_{\nu}} < 1.1$ dominate most of the area of the parameter space. The area where ${\cal R}_{\nu} > 1.1$ only occurs where the disk is optically thin and the albedo is large. This increase occurs due to a change in the optical depths regimes in the absorption and scattering cases. In this region $\tau_{\kappa_{\nu}}< 1$ but $\tau_{\chi_\nu} > 1$; then, the emergent intensities can be approximated as $I_{\nu}^{\rm abs} \approx B_{\nu}(T)\tau_{\kappa_{\nu}}$, and $I_{\nu}^{\rm sca} \approx B_{\nu}(T)$,  respectively, and $R_{\nu} \sim 1/\tau_{\kappa_{\nu}} > 1$. The region where ${\cal R}_{\nu} < 0.9$ occurs in the optically thick regime and high albedo. The latter increases the total optical depth of the disk and the mean free path of the photons decreases. In consequence, a small fraction of photons can escape from the disk and the emergent intensity $I_{\nu} ^{\rm sca}$ decreases. For example, for $\omega_{\nu} = 0.95$, the ratio ${\cal R}_{\nu} = 0.44$.

\section{Spectral index}
\label{SEC:spectral_index}
Even if the albedo is almost constant for mm and cm dust grains, the change of the optical depth at different wavelengths modifies the emergent intensity $I_{\nu}^{\rm sca}$ compared with the absorption case, as shown in Figure (\ref{FIG:Ratio_Intensities}). If the albedo is $\omega_{\nu} \gtrsim$ 0.6, the emergent intensity in the scattering case decreases at optically thick wavelengths and increases at $-2 \lesssim \log(\tau_{\kappa_{\nu}}) \lesssim -1$. Then, one expects changes of the spectral indices in the scattering case with respect to the pure absorption case.

The spectral index between the frequencies $\nu_1$ and $\nu_2$ in the pure absorption case is defined as
\begin{equation}
\alpha_{\nu_1, \nu_2}^{\rm abs} =  \frac{\log \left( I_{\nu_1}^{\rm abs} / I_{\nu_2}^{\rm abs} \right)}{\log \left( \nu_1/\nu_2 \right)},
\end{equation}
while in the scattering case the spectral index is
\begin{equation}
\alpha_{\nu_1, \nu_2}^{\rm sca} =  \frac{\log \left( I_{\nu_1}^{\rm sca} / I_{\nu_2}^{\rm sca} \right)}{\log \left( \nu_1/\nu_2 \right)}.
\end{equation}

From the definition of ${\cal R}_{\nu}$ (equation \ref{EQ:R}) the spectral indices are related by
\begin{equation}
\alpha_{\nu_1,\nu_2}^{\rm sca} = \frac{\log \left( {\cal R}_{\nu_1}/ {\cal R}_{\nu_2} \right)}{\log \left( \nu_1/\nu_2 \right)} + \alpha_{\nu_1, \nu_2}^{\rm abs}.
\label{EQ:spectral_index_relation}
\end{equation}

\begin{figure}
\centering
\includegraphics[scale=0.65]{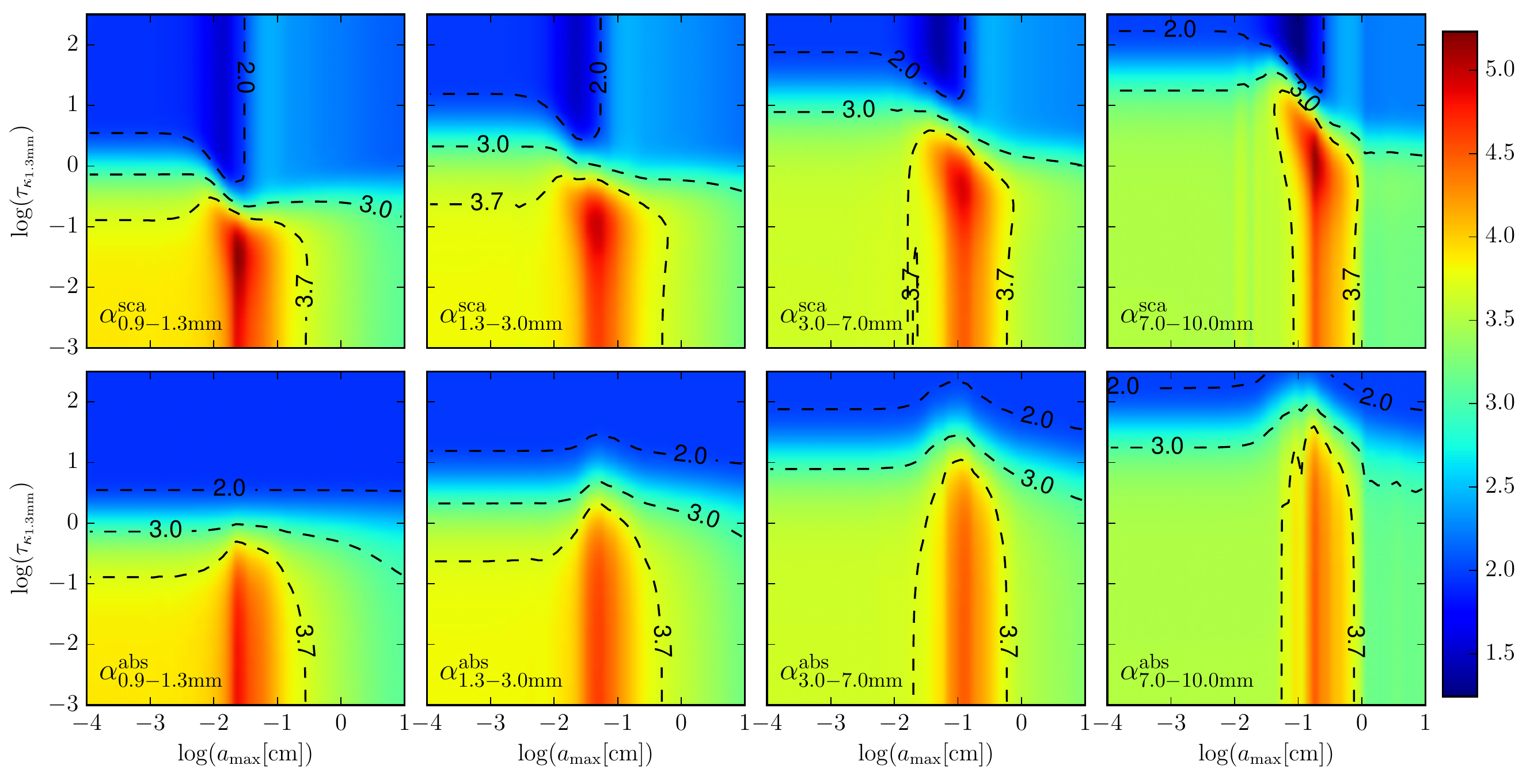}
\caption{Spectral indices in the mm range as a function of the optical depth at 1.3 mm and the maximum grain size. The slope of the particle size distribution if fixed to $p=3.5$ and the temperature is $T = 100$ K. In the top panels the scattering effects are taken into account, while in the bottom panels they are ignored. The color bar is the same in all panels.}
\label{FIG:spectral_index}
\end{figure}

Top panels of Figure (\ref{FIG:spectral_index}) show the spectral indices $\alpha^{\rm sca}_{\lambda_1,\lambda_2}$\footnote{For simplicity we change the notation from $\alpha^{\rm sca}_{\nu_1,\nu_2}$ to $\alpha^{\rm sca}_{\lambda_1,\lambda_2}$, where $\lambda \nu = c$ and $c$ is the speed of light.} between the consecutive wavelengths $\lambda = 0.87, 1.3, 3.0, 7.0$ mm and 1 cm, from left to right as a function of the the maximum grain size and the absorption optical depth at $\lambda = 1.3$ mm (from now on, this optical depth is taken as the reference to define the optical depth regime when computing the spectral indices). The slope of the particle size distribution is set to $p=3.5$ and the temperature is set to $T = 100$ K. 
These indices change in the case of  very cold disks ($T \lesssim $30  K), where the peak of the black body radiation is displaced to the sub-mm range according to Wien's law. This is shown in Figure (\ref{FIG:Spectral_index_cold}) in Appendix (\ref{App:RJeans}), where the spectral indices are computed at $T = 10$ K.
 The bottom panels of the same Figure show the spectral indices in the pure absorption case $\alpha^{\rm abs}_{\lambda_1,\lambda_2}$. The color bar is the same for all panels. Isocontours where the spectral index is $2.0$ (typical of optically thick emission in Rayleigh-Jeans limit), 3.0, and 3.7 (the ISM value), are shown in all the panels as black dashed lines.

In the absorption case (bottom panels) one can see that for very optically thick disks ($\log(\tau_{\kappa_{1.3\rm{mm}}}) >> 0$), the value of $\alpha^{\rm{abs}}_{\lambda_1,\lambda_2}$ tends to 2. Usually, values of $\alpha < 3.7$ are interpreted as grain growth for disks in the optically thin regime (e.g. \citealt{Beckwith_1991}, \citealt{Natta_2004}). See, for example, that the spectral index $\alpha_{7.0-10.0 \rm{mm}}^{\rm{abs}}$ is between $3.0$ and $3.7$ for optically thin disks ($\log(\tau_{\kappa_{1 .3\rm{mm}}}) < 0$) and large dust grains ($\log(a_{\rm{max}}[\rm{cm}] )\gtrsim  -0.5$). However, these spectral indices can also be explained by disks in the optically thick regime ($ 0.5 \lesssim \log (\tau_{\kappa_{1.3 \rm{mm}}}) \lesssim 1.5$) and very small grains ($\log(a_{\rm{max}}[\rm{cm}] )\lesssim  -2$).

The scattering case (top panels) in the optically thin region has the same spectral indices than in the pure absorption case.  This occurs because the emergent intensities with or without scattering coincides at this optical depth regime, ${\cal R}_{\nu} \sim 1$ (equation \ref{EQ:Optical_thick_limit}). This behaviour is evident in Figure (\ref{FIG:spectral_index}) for $\log(\tau_{\kappa_{1.3 \rm{mm}}}) < -2$. 

In  the optically thick regime ($\log(\tau_{\kappa_{1.3\rm{mm}}}) >> 0$), there are two regions in the parameter space: the first region has $\log(a_{\rm max} [\rm{cm}])< -2$, which has an albedo around $0$ at all wavelengths (see Figure 4 of \citealt{Carrasco_2019}), therefore ${\cal R}_{\nu} \sim 1$ (see Figure \ref{FIG:Ratio_Intensities}), and one has the same spectral index than in the pure absorption case. The second region has $\log(a_{\rm max} [\rm{cm}])> -2)$, which has a significant albedo and where the emergent intensities (compared with the pure absorption case) change according to the albedo and optical depth (Figure \ref{FIG:Ratio_Intensities}). Any change of the albedo or  the optical depth with the frequency modifies ${\cal R}_{\nu}$ and, thus, the spectral index in the scattering case changes compared to the true absorption case according to equation (\ref{EQ:spectral_index_relation}).

For example, if a disk is optically thick at 3 mm but optically thin at 7 mm, the emergent intensity decreases and increases respectively compared with the true absorption case. Therefore, the spectral indices between these wavelengths in the scattering and true absorption cases do not coincide.

The spectral indices maps in Figures (\ref{FIG:spectral_index}) and (\ref{FIG:Spectral_index_cold}) can be used to give an idea of the values of $a_{\rm max}$ and $\tau_{\kappa_{1.3mm}}$ which can produce an observed spectral index. On the other hand, multi-wavelength observations (more than 2 frequencies) allow the determination of the albedo, the optical depth, and the disk temperature by fitting equation (\ref{EQ:Intensity_scattering}).
Given the albedo, one can then find the corresponding value of $a_{\rm max}$.
 
\begin{figure}
\centering
\includegraphics[scale=0.85]{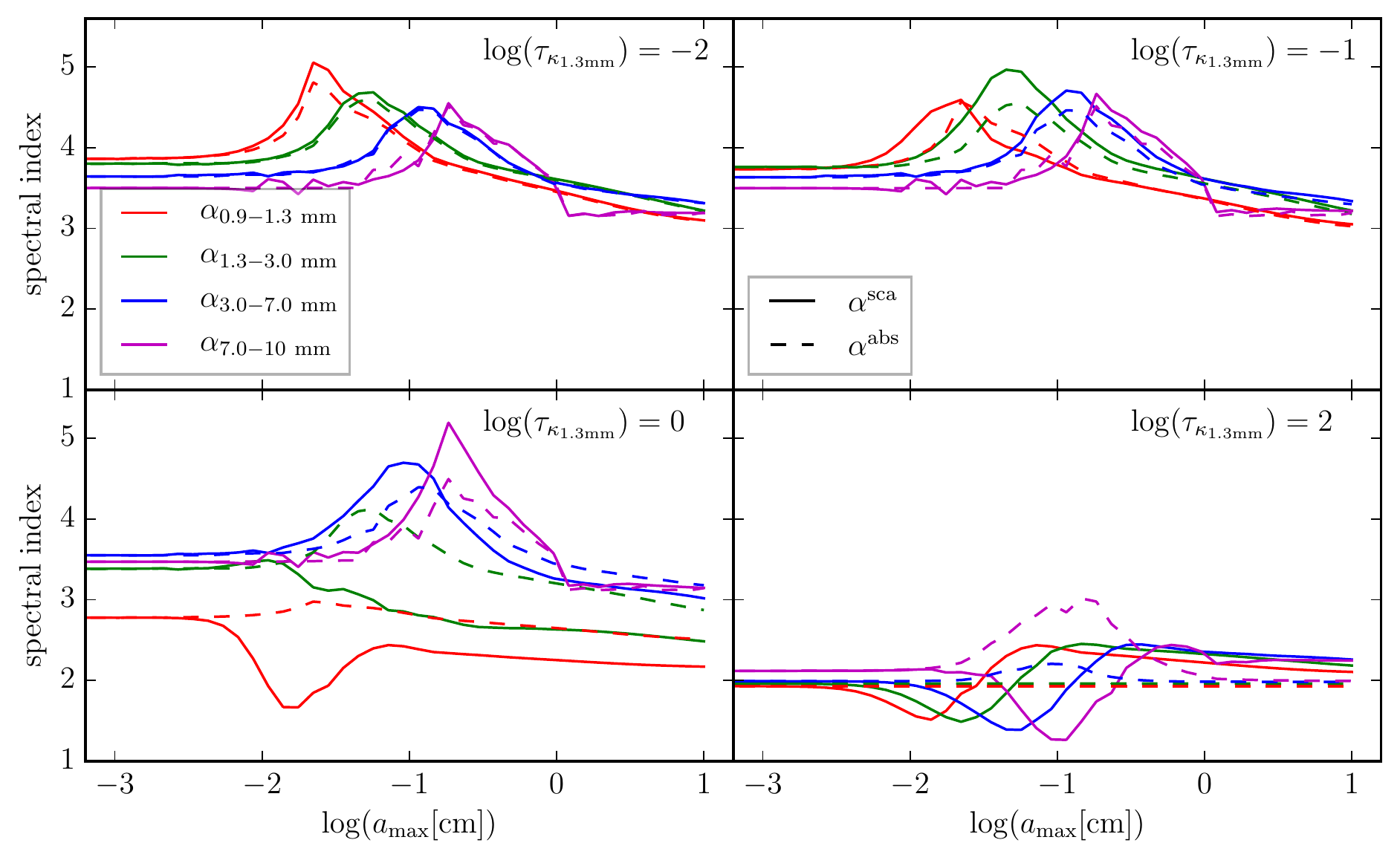}
\caption{Spectral indices as a function of $a_{\rm max}$ for $\log(\tau_{\kappa_{1.3 \rm{mm}}}) = -2, -1, 0$ and 2 from left to right and from top to bottom respectively. The solid and dotted curves are the scattering and the pure absorption models respectively, while each color represent the same spectral index summarized in the top left panel.}
\label{FIG:spectral_index_corte}
\end{figure}

Figure (\ref{FIG:spectral_index_corte}) shows slices of the spectral indices in Figure (\ref{FIG:spectral_index}) at constant optical depth. These indices coincide with those shown in Figure 9 of \cite{Zhu_2019} in the optically thin ($\tau_{\kappa_{1.3 \rm{mm}}} = -2$, top left panel) and the optically thick regime ($\tau_{\kappa_{1.3 \rm{mm}}} = 2$ bottom right panel). Slices at intermediate optical depths ($\tau_{\kappa_{1.3 \rm{mm}}} = -1,0$ are also shown in the top right and bottom left panels, respectively).  

In all the cases, the spectral index in the scattering and pure absorption cases coincide for $\log({a_{\rm max}/{\rm cm}}) \lesssim -2.5$, where $\omega_{\nu} \sim 0$.  Also, they are the same for all the grain sizes in the optically thin regime at all frequencies (top right panel). Furthermore, in this limit, the value of the spectral index coincides with the typical assumption $\alpha^{\rm abs} = \alpha^{\rm sca} = \beta_{\kappa_{\nu}} + 2$.

The bottom right panel of Figure (\ref{FIG:spectral_index_corte}), which is the most optically thick case, the spectral indices can be lower than $2$ in the scattering case. For example, the values of the maximum grain size where  $\alpha_{0.9-1.3 \rm mm} < 2$ vary from $ 10^{-2.5} \lesssim a_{\rm max}/\rm{cm} \lesssim10^{-1.5}$, consistent with the discussion in \cite{Liu_2019}. 
Note that larger grains are required to produce spectral indices smaller than 2 at longer wavelengths.

Recently, \cite{Li_2017} and \cite{Galvan_2018} proposed that these low spectral indices can also be produced in the case of pure absorption emission by an optically thick disk with a radially decreasing temperature and self-obscuration. We will explore the temperature effects in the optically thin and thick regime and in the pure absorption and scattering cases in Section \ref{SEC:Temperature}.

The solution to the radiative transfer equation with scattering (equation \ref{EQ:Intensity_scattering}) for vertically isothermal disks was used in \cite{Carrasco_2019} to fit the dust temperature, dust surface density, and  maximum grain size in the HL Tau disk. In their Appendix C, they showed the difference between the inferred dust parameters in the scattering and true absorption cases. In this source, the inferred values of $a_{\rm max}$ are smaller by a factor of 10 when scattering is included in the radiative transfer equation.
This solution was also implemented in \cite{Sierra_2019} to fit the dust continuum observations in the disk around HD 169142.
They found that the maximum grain size at the center of the inner disk ring  at $\sim 27$ au varies from 10 cm to 2 mm, depending on the assumed value of the slope of the particle size distribution (see their Figure 6).

Other authors have also studied the inclusion of the scattering effects in the radiative transfer.  
For example, \cite{Birnstiel_2018} found an analytic solution for vertically isothermal slabs with the Eddington-Barbier approximation. Using their solution, \cite{Liu_2019} found spectral indices below of 2, consistent with the results discussed in this section. 
In addition,  \cite{Soon_2017} integrated the radiative transfer equation by ray tracing includying scattering, to model the $\lambda = 870 \ \mu$m observations of the disk around HD 142527  and inferred its physical properties.
 
In this section, to derived the spectral indices one assumed a vertically isothermal disk and a constant maximum grain size.
 However, it is known that the disk temperature varies as a function of the height above the disk mid plane because the stellar irradiation (which heats the disk surface) does not penetrate inside optically thick mid plane regions,  where the temperature is determined by viscous heating \citep{Dalessio_1998}. Also, dust settling has been observationally inferred in protoplanetary disks (e.g. \citealt{Pinte_2016}), where the largest grains concentrate around the mid plane (e.g. \citealt{Dubrulle_1995}). 
 
Sections (\ref{SEC:Temperature}) and (\ref{SEC:settling}) explore the effect of a vertical temperature gradient and dust settling on the spectral indices.
The analytical solution of the emergent intensity (equation \ref{EQ:Intensity_scattering}) is no longer used in these sections, since it assumes a vertically isothermal well-mixed disk. Then, the radiative transfer equation (equation \ref{EQ:RadTrans}) is numerically integrated.

\newpage
\section{Temperature effects}
\label{SEC:Temperature}
This section explores the effects of a vertical temperature gradient on the emergent intensity and the spectral indices.
The maximum grain size as a function of the height above the mid plane is assumed constant (e.g. no settling effects are taken into account).

We consider a simple temperature model where the disk is heated by viscous dissipation (accretion) and stellar irradiation. The first one heats the disk mid plane and the second one heats the disk surface. We follow the results from \cite{Calvet_1991} to compute the structure of the vertical temperature. In that work, the temperature due to the accretion and irradiation by the central star $T(z)$ is given by 
\begin{equation}
T^4(z) =  T^4_{\rm acc}(z) +  T^4_{\rm irr}(z),
\label{EQ:Temp4}
\end{equation}
where
\begin{equation}
T^4_{\rm acc}(z) = \frac{3}{4} \left( \tau_{\rm R} + \frac{2}{3} \right)  T^4_{\rm eff}
\end{equation}
is the disk temperature due to accretion (it is maximum at the disk mid plane), $\tau_{\rm R}$ is the Rosseland optical depth measured perpendicular to the disk mid plane, and
\begin{equation}
T_{\rm eff}^4 =  \frac{3 \dot{M} \Omega_{\rm K}^2}{8\pi \sigma_{\rm B}}
\label{EQ:Teff}
\end{equation}
is the effective temperature of a disk with keplerian angular velocity $\Omega_{\rm K}$ and accretion rate $\dot{M}$, and where $\sigma_{\rm B}$ is the Stefan-Boltzmann constant.

In addition, the temperature due to stellar irradiation is
\begin{equation}
T^4_{\rm irr}(z) =  \frac{L_*}{16\pi \varpi^2 \sigma_{\rm B}} \exp\left(-\frac{q\tau_{\rm R}}{\varphi}\right)  =  \frac{T_*^4}{4} \left(\frac{R_*}{\varpi} \right)^2 \exp\left(-\frac{q\tau_{\rm R}}{\varphi}\right),
\end{equation}
where $L_*$, $T_*$ and $R_*$ are the luminosity, temperature, and radius of the central star. The angle between the stellar irradiation and the disk surface is $\varphi$ (typical value of $\varphi = 0.05$), and $q$ is the ratio between the stellar and the disk Rosseland optical depths.

\begin{figure}
\centering
\includegraphics[scale=0.8]{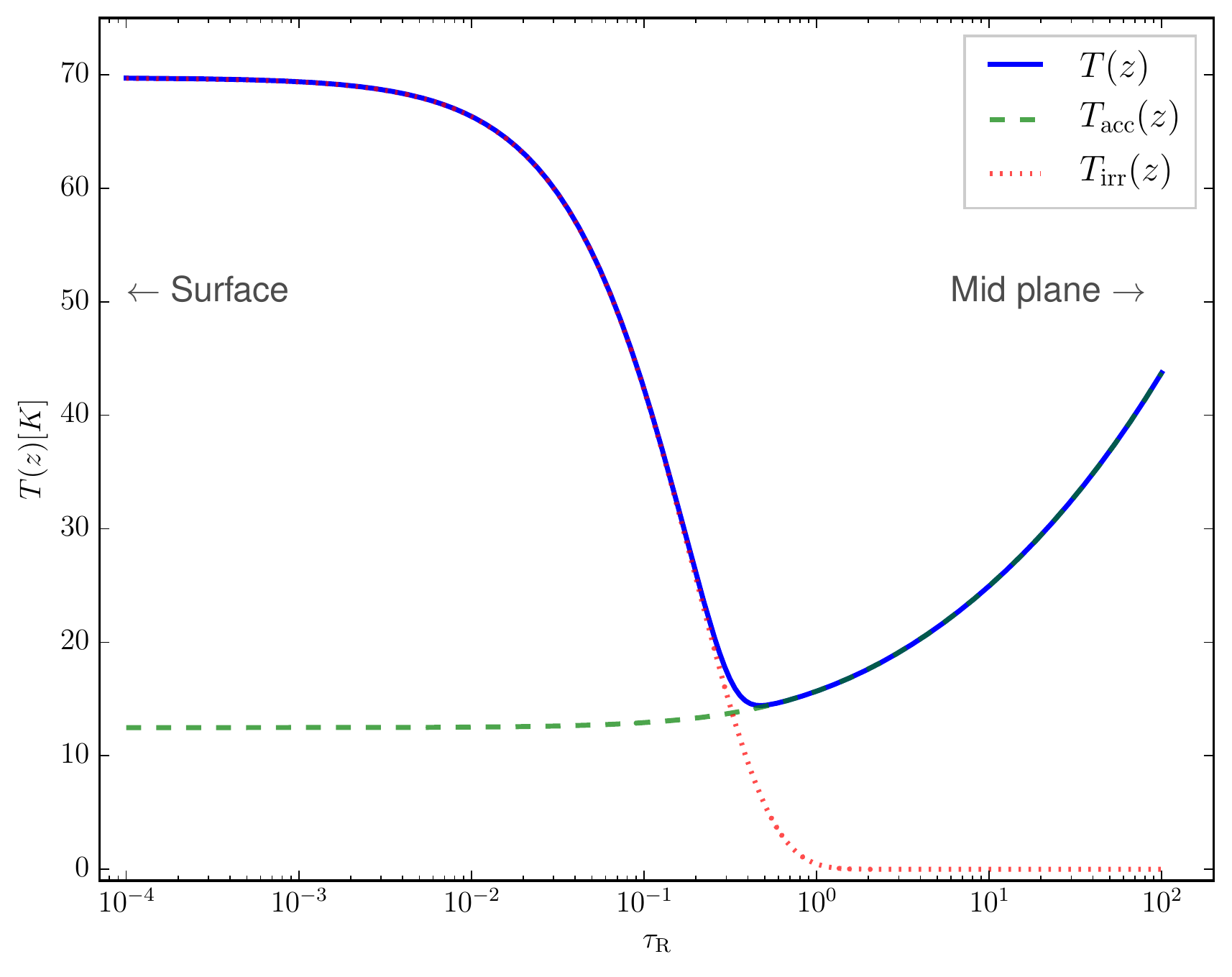}
\caption{Disk temperature above the mid plane as a function of the Rosseland opacity. The red dotted line is the irradiation temperature, the green dashed line is the accretion temperature, and the blue solid line is the total temperature profile. The Rosseland optical depth is measured from the disk surface, such that the mid plane and surface of the disk are at the right and left of this plot, respectively.}
\label{fig:Temperature_opt}
\end{figure}

From now we assume a representative case $q=1$ \citep{Calvet_1991}, and the Rosseland opacities are computed using the results from Table 1 of \cite{Dalessio_2001}, which are a factor of $\sim 7$ larger than the extinction coefficient at $\lambda = 1.3$ mm for grains of 1 cm. The temperature profile obtained from equations (\ref{EQ:Temp4}-\ref{EQ:Teff}) is shown in Figure (\ref{fig:Temperature_opt}) for the following parameters: $\dot{M} = 3 \times 10^{-8} \ M_{\odot}$ yr$^{-1}$, $M_* = 0.3 \ M_{\odot}$, $R_* = 1.7 \ R_{\odot}$, $L_* = 0.38 \ L_{\odot}$. 
These parameters correspond to a typical T Tauri star with an age of 1.5 Myr (Manzo et al. submitted). We calculate the temperature profile at $\varpi = 10$ au. We remark that we are not trying to give a full model of the vertical temperature gradient, but only a realistic profile which can be used to compute their effects on the emergent emission and spectral indices.
The red dotted line is the irradiation temperature, the green dashed line is the accretion temperature, and the blue solid line is the total temperature profile (equation \ref{EQ:Temp4}). This profile is similar to those shown for example in Figure 4 of \cite{Dalessio_1998} or Figure 2 of \cite{Calvet_1991}. 

\begin{figure}
\centering
\includegraphics[scale=0.55]{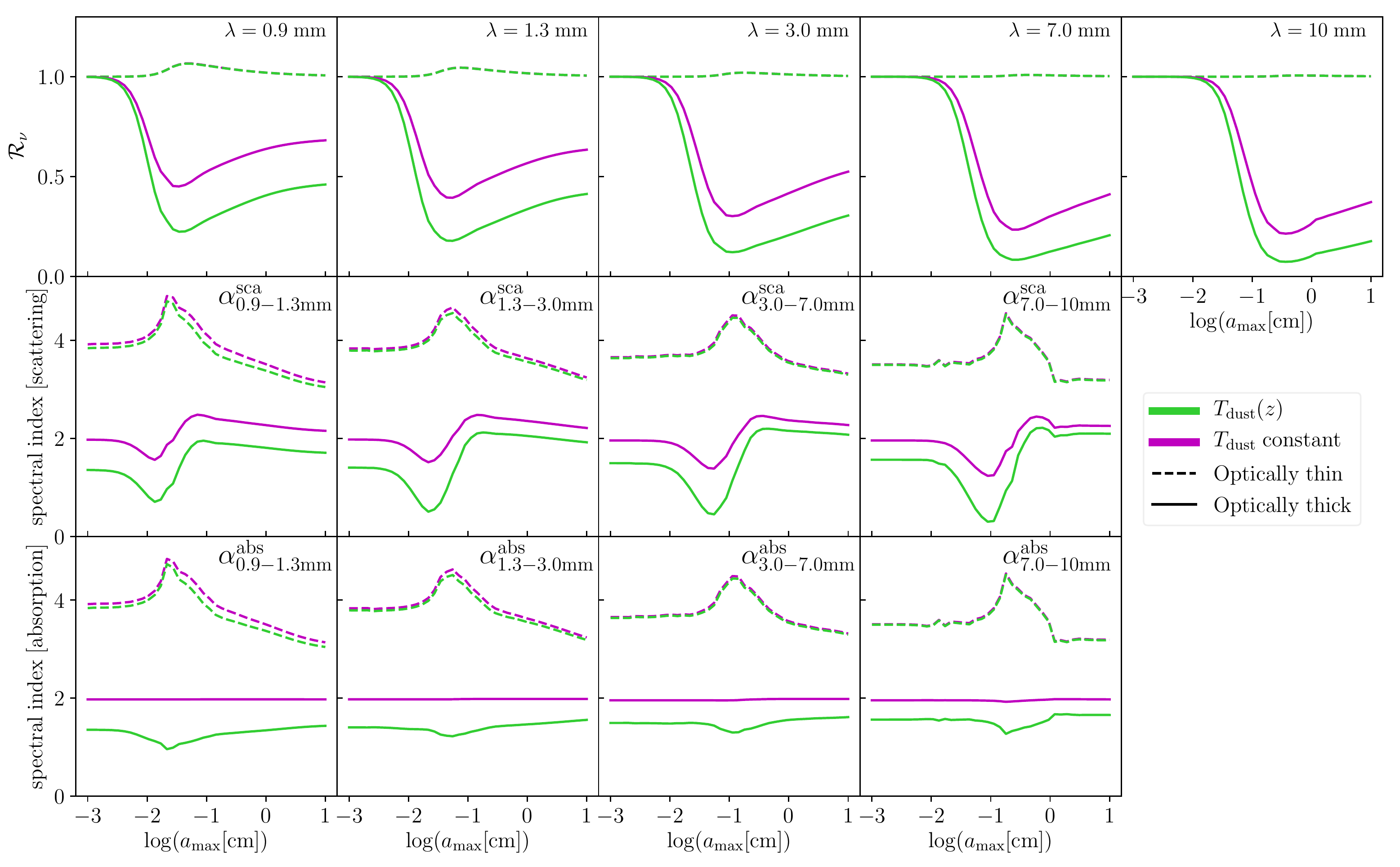}
\caption{Effect of the temperature gradient in the ratio ${\cal R}_{\nu}$ (first row), the spectral index in the scattering and absorption case (second and third row, respectively) as a function of the maximum grain size. The wavelength increases from left (0.9 mm) to right (1 cm). In all the panels, the green lines are the properties with a temperature model that varies with the height above the mid plane and the magenta lines have a constant temperature. The dashed and solid lines are the optically thin and optically thick regimes, respectively.}
\label{fig:Temperature}
\end{figure}

Figure (\ref{fig:Temperature}) shows the effects of the vertical temperature gradient on the ratio ${\cal R}_{\nu}$ (first row), the spectral index in the scattering case (second row) and in the absorption case (third row). The wavelength is indicated in the top right corner of each panel. In all the panels, the green and magenta lines represent the properties with and without vertical temperature gradient, respectively; and the dashed and solid lines are the solution for the optically thin ($\log (\tau_{\kappa_{1.3} \rm mm}) << 0 $) and thick regimes ($\log (\tau_{\kappa_{1.3} \rm mm}) >> 0 $), respectively, which is the same definition as in the previous section based on the absorption optical depth at $\lambda = 1.3$ mm.

When the vertical temperature structure is taken into account, the ratio ${\cal R}_{\nu}$ does not change in the optically thin regime, but it decreases compared with the vertically isothermal model in the optically thick regime. The latter occurs because the height above the mid plane where the disk becomes optically thick is larger in the scattering case compared with that of the pure absorption case, then, the emergent intensity in the pure absorption case has a larger influence of the hot region close to the mid plane.

The spectral indices for constant temperature coincide with the results shown in Figure (\ref{FIG:spectral_index_corte}). In the optically thin regime, the spectral indices do not change when the temperature gradient is taken into account. This occurs because there are not hidden zones of the disk and all the disk contributes to the emergent intensity. 
In the optically thick regime, the spectral indices when the temperature structure is taken into account are smaller than those inferred with a vertically isothermal disk. This occurs because the larger wavelengths can penetrate to regions closer to the mid plane, where the temperature rapidly increases due to viscous heating, so one expects that the emergent intensity at the larger wavelength increases relative to that of the smaller wavelength, which decreases the spectral indices compared with the vertically isothermal model. For example, the spectral index between 7 mm and 1 cm could reach a value close to 0 in the optically thick case if the maximum grain size is $\sim 1$ mm.

The spectral indices in the absorption case show the well known properties, they are $\alpha^{\rm abs} = \beta_{\kappa_{\nu}} + 2$ in the optically thin regime, $\alpha^{\rm abs} = 2$ in the optically thick regime with constant temperature, and $\alpha^{\rm abs} < 2$ in the optically thick regime with a gradient temperature that increases toward the mid plane (the latter behavior was also found by \citealt{Li_2017} and \citealt{Galvan_2018}).

\section{Dust settling}
\label{SEC:settling}
This section considers the differential change of the dust maximum grain size as a function of the height above the mid plane due to dust settling and its effect on the dust emission and spectral indices. The mechanism that prevents a perfect settling of all the dust grains in the mid plane is turbulence. Small grains are well coupled to the gas and they are expected to be found in all the disk. Large grains are less coupled to the gas, thus, these grains tend to accumulate around the mid plane due to the vertical component of the stellar gravity, so one can expect that they are depleted from the upper layers of the disks.

\cite{Dubrulle_1995} showed that the scale height of dust grains with Stokes number St is given by 
\begin{equation}
h_{\rm d}(a) = H_{\rm g} \left[  1+ (\gamma + 1)^{1/2} \frac{\rm St}{\alpha_t}    \right]^{-1/2},
\label{eq:dust_scale_height}
\end{equation}
where $H_{\rm g}$ is the gas scale height, $\gamma$ is an exponent related to the scale where the energy is injected ($\gamma$ = 2 for compressible turbulence), $\alpha_t$ is the turbulence parameter given by the \cite{Shakura_1973} equation, and ${\rm St} = \pi \rho_{\rm m} a/(2 \Sigma_{\rm g})$ is the Stokes number for a grain with material density $\rho_{\rm m}$ and within a medium with a gas surface density $\Sigma_{\rm g}$. The dust density for grains of size $a$ is given by
\begin{equation}
\rho_{\rm d}(a,z) = c_1 \exp \left[ - \frac{1}{2} \frac{z^2}{h_{\rm d}(a)^2} \right],
\end{equation}
where $c_1$ is a coefficient that depends on the total dust density in the mid plane (see derivation below).
If one considers that the dust grains with size $a$ are completely depleted at a height $z = 3 h_{\rm d}(a)$ (where their density decreases by a factor of 99.7\% with respect with the mid plane density), then, from equation (\ref{eq:dust_scale_height}) one can derive the maximum grain size as a function of the height above the mid plane as
\begin{equation}
a_{\rm max}^{\rm settling}(z) = \frac{2 \Sigma_{\rm g} \alpha_t}{\pi \rho_{\rm m}} (\gamma+1)^{-1/2} \left[\frac{9H_{\rm g}^2}{z^2} -1 \right] .
\end{equation}
This equation gives the limit of the maximum grain size at a height $z$ due to turbulent mixing. The maximum grain size is also limited by the largest grain size in the disk. Then, if the largest grain size in the disk is 1 cm, the maximum grain size is given by 
\begin{equation}
a_{\rm max}(z)  = \min (a_{\rm max}^{\rm settling}(z), \  \rm{1 \ cm}).
\label{eq:Max_grain_size}
\end{equation}

In addition, dust settling changes the particle size distribution as a function of the height above the mid plane. The particle size distribution after dust settling $n_{\rm sett}(a)$ is given by (see equation 32 of \cite{Sierra_2019})
\begin{equation}
n_{\rm sett}(a,z) \propto n(a) \frac{\rho_{\rm d}(a,z)}{h_{\rm d}(a)},
\label{eq:new_particle_size_dist}
\end{equation}
where $n(a)$ is the original particle size distribution. Note that, for large values of $\alpha_t$, the dust scale height $h_{\rm d}(a)$ and the dust density $\rho_{\rm d}(a,z)$ become independent of the grain size and the particle size distribution does not change. The right hand of equation (\ref{eq:new_particle_size_dist}) is fitted using a power law $n_{\rm sett}(a) \propto a^{-p_{\rm sett}}$ between a minimum grain size $a_{\rm min} = 0.05\ \mu$m and the maximum grain size $a_{\rm max}(z)$ in order to obtain a particle size distribution and the local opacity properties.

If one assumes $n(a)da \propto a^{-3.5}da$ and $a_{\rm max} >> a_{\rm min}$, one can obtain an analytic solution for the total dust density that takes into account the contribution of all the grain sizes as\footnote{This definition is equivalent to equation 10 of \cite{Sierra_2019}, where the dust redistribution (settling) is taken into account by the factor $\rho_{\rm d} (a,z)$, and the integral can be computed using the global values of the disk $a_{\rm max} = 1$ cm and $p = 3.5$.}
\begin{equation}
 \rho_{\rm d}(z) = \frac{\int_{a_{\rm min}}^{a_{\rm max}} \rho_{\rm d} (a,z) a^3 n(a)da }{ \int_{a_{\rm min}}^{a_{\rm max}} a^3 n(a)da } = 
 \frac{c_1}{2} \sqrt{\pi} \exp \left[ - \frac{1}{2} \frac{z^2}{H_{\rm g}^2} \right] {\erf} \left[ \frac{z}{\sqrt{2}H_{\rm g}} \sqrt{\frac{a_{\rm max}}{a_0}}   \right] \left( \frac{z}{\sqrt{2}H_{\rm g}} \sqrt{\frac{a_{\rm max}}{a_0}} \right)^{-1},
\label{eq:settled_density}
\end{equation}
where $a_0 = 2 \Sigma_{\rm g} \alpha_t (\gamma+1)^{-1/2}/(\pi \rho_{\rm m})$ and $\rm erf$ is the error function. For a large value of the turbulent parameter $\alpha_t$, the argument within the error function tends to zero. In this limit, ${\erf}(x)/x \rightarrow 2/\sqrt{\pi} + {\cal O} (x^2)$, and the dust density is given by $\rho_{\rm d}(z) = c_1 \exp \left[ -z^2/(2H_{\rm g}^2) \right]$, as expected for a well-mixed disk. 
By mass conservation, the column density of a well-mixed disk has the same column density as a settled disk (derived from equation \ref{eq:settled_density}). Therefore, from the condition $\int _{-\infty} ^{\infty} \rho_{\rm d}^{\rm mix}(z)dz = \int _{-\infty} ^{\infty} \rho_{\rm d}(z)dz$,
one obtains
\begin{equation}
c_1 = \rho_{\rm d,0} \left[ \frac{\sqrt{\frac{a_{\rm max}}{a_0}}}{ \sinh^{-1} \left( \sqrt{\frac{a_{\rm max}}{a_0}} \right) } \right].
\end{equation}

\begin{figure}
\centering
\includegraphics[scale=0.55]{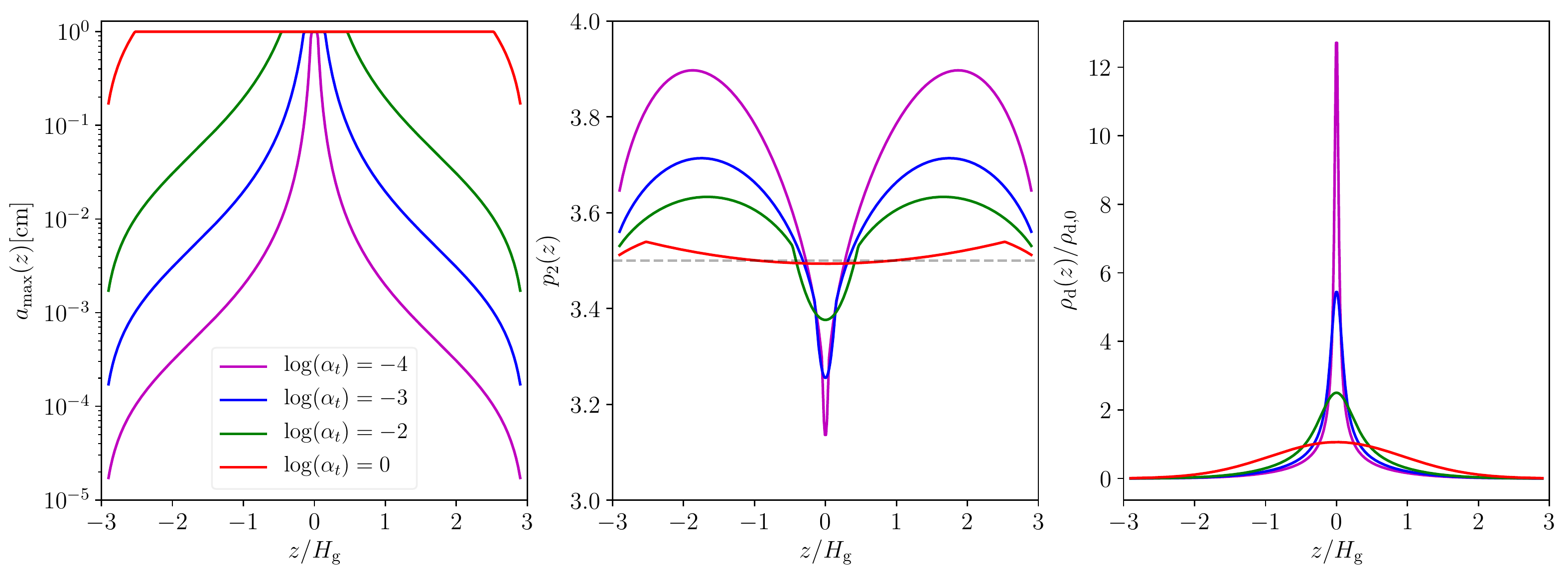}
\caption{Settling model. Maximum grain size (left panel), slope of the particle size distribution (middle panel), and dust density (right panel) as a function of the height above the mid plane. The color of each curve corresponds to a model with different $\alpha_t$ parameter, see legend at the left panel.}
\label{FIG:settling}
\end{figure}

For a large turbulent parameter, $x/\sinh^{-1}(x) \rightarrow 1$ and $c_1 \rightarrow \rho_{\rm d,0}$ as expected. Figure (\ref{FIG:settling}) shows the maximum grain size (left panel) and the slope of the particle size distribution (middle panel) and the total dust density (right panel) as a function of the height above the mid plane for different $\alpha_t$ values and assuming that the maximum grain size is 1 cm. The value of $\log (\alpha_t) = 0.0$ (high turbulence, red curve) recovers the same maximum grain size at all height above the mid plane, a constant slope of $p \approx 3.5$, and a gaussian dust density with scale height $H_{\rm g}$ (the same than that of the gas). However, typical values of the $\alpha_t$ parameter ($\sim 10^{-3}$) are not high enough to prevent the dust settling, e.g. the maximum grain size is around $30\ \mu$m, $200\ \mu$m and $2$ mm at $z/H_{\rm g} = 1$ for $\log(\alpha_t) = -4,-3,-2$, respectively, while the 1 cm grains are only around the mid plane. The slope $p< 3.5$ close to the mid plane and $p>3.5$ for $z/H_{\rm g} \gtrsim 0.4$. In addition, the dust density is enhanced around the mid plane due to dust settling. The magnitude of the density at the mid plane increases by a factor of $\sim 12, 5, 2$ compared with the well-mixed disk for $\log(\alpha_t) = -4,-3,-2$, respectively.

\begin{figure}
\centering
\includegraphics[scale=0.55]{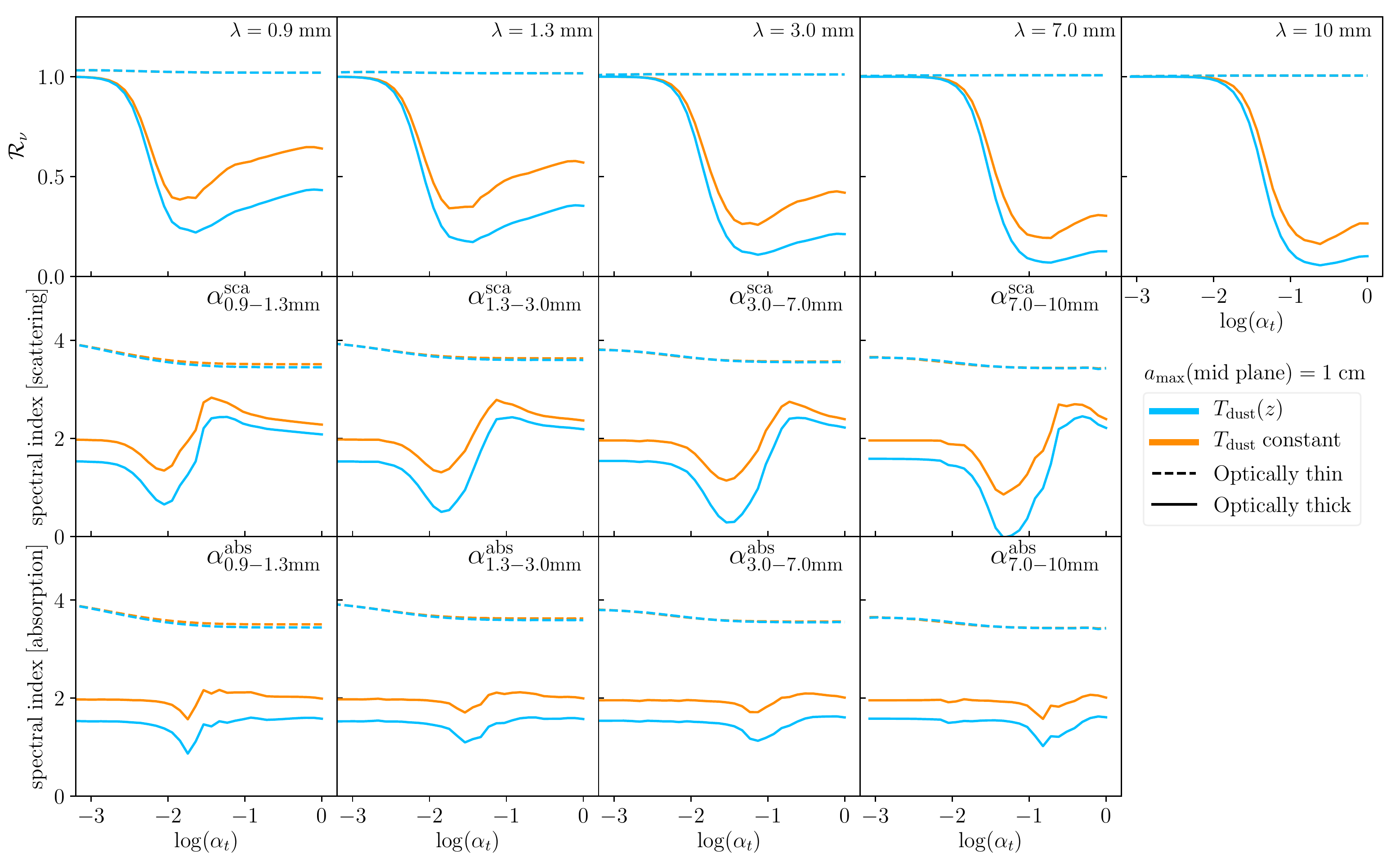}
\caption{Effect of the settling in the ratio ${\cal R}_{\nu}$ (first row), the spectral index in the scattering and absorption case (second and third row, respectively) as a function of the $\alpha_t$ parameter. The wavelength increases from left (0.9 mm) to right (1 cm). In all the panels, the light blue lines are the properties with a temperature model that varies with the height above the mid plane and the orange lines have constant temperature. The dashed and solid lines are the optically thin and optically thick regimes, respectively.}
\label{fig:alpha_settling}
\end{figure}

Figure (\ref{fig:alpha_settling}) shows the effects of the dust settling on the ratio ${\cal R}_{\nu}$ (first row), the spectral indices in the scattering case (second row), and in the absorption case (third row) as a function of the $\alpha_t$ parameter. The wavelength is indicated in the top right corner of each panel. In all the panels, the light blue and orange lines represent the properties with and without vertical temperature gradient, respectively; and the dashed and solid lines are the optically thin ($\log (\tau_{\kappa_{1.3} \rm mm}) << 0 $) and thick regimes ($\log (\tau_{\kappa_{1.3} \rm mm}) >> 0 $), respectively. Similarly to the results of Figure (\ref{fig:Temperature}), the temperature vertical gradient decreases all the spectral indices compared with the vertically isothermal model.

The maximum grain size in all the models is 1 cm, however, they settle around the mid plane with a scale height that depends on the turbulence parameter (Figure \ref{FIG:settling}). Models with small values of $\alpha_t$ have all their centimeter grains within a small region around the mid plane. Thus, the small grains in the upper layers can hide the emission of the large grains and change the spectral indices if the disk is optically thick. Models with large $\alpha_t$ have all the dust grains well-mixed in the disks, then, the spectral indices corresponds to the properties of 1 cm grains.

The similarity between Figure (\ref{fig:alpha_settling}) and Figure (\ref{fig:Temperature}) in the optically thick regime is not a coincidence. For small turbulence $(\log (\alpha_t) < -2)$ and in the optically thick regime, the disk emission comes from the small grains in the upper layers, which have a small albedo at mm wavelengths, then $\cal{R}_{\nu} \sim$ 1 and $\alpha^{\rm sca} = \alpha^{\rm abs} = 2$. Note from equation (\ref{eq:Max_grain_size}) that the maximum grain size in the disk surface increases with $\alpha_t$, then, if the disk is optically thick, the spectral indices are mainly tracing the properties of the grains in the disk surface.

In the optically thin regime, there are not hidden dust grains due to optical depth effects, thus, ${\cal{R}_{\nu}} \sim 1$ for all $\alpha_t$, and the spectral indices correspond to the properties of the 1 cm grains independent of the degree of settling.

\subsection{Maximum grain size inferred from optically thick and settled disks}
\label{SUBSEC:amax_settling}
As discussed in the Introduction, it has been proposed that millimeter polarization observations only trace the small 
dust in the upper layers of a settled disk \citep{Yang_2017}.
However, the large dust grains (millimeter, centimeter sized) could be hidden in a settled disks only if the disk becomes optically thick at a height where the is no emission from large grains. This means that one requires that the disk's upper layers, with dust grains of hundred microns sizes, are optically thick.

There are two main problems for such a scenario: The opacity is mainly dominated by the mm-cm grain size in the disk. The opacity of some hundred micrometer grains (or smaller) at millimeter wavelengths is one or two orders of magnitude smaller than the mm-cm grains (see Table 1 of \cite{Dalessio_2001} and Figure 10 of \cite{Sierra_2017}). Also, the dust density decreases with the height above the mid plane, then, the available mass in the upper layers of the disk  surface is small compared with that close to the mid plane, making it difficult for the disks to become optically thick at their surface. The total dust column density has to be very high in order to compensate for the two effects.

The minimum dust column density needed can be computed as follows:
Suppose that the disk is settled with $\log (\alpha_t) = -3$, then, all the grains above $z = H_{\rm g}$ are $\sim 100 \ \mu$m or smaller (see Figure \ref{FIG:settling}). For a vertically isothermal disk, the column density above a height $z$ is given by 
\begin{equation}
\Sigma^{\rm up}_{\rm d} = \frac{\Sigma_{\rm d} }{2} \left[ 1 - \rm{erf} \left( \frac{z}{\sqrt{2} H_{\rm g}}\right)  \right],
\end{equation}
For $z = H_{\rm g}$, the factor within the brackets is $\approx 0.317$. Then, the column density above one gas scale height is around $\sim 1/6$ of the total column density. If one requires that the disks becomes optically thick at $z >  H_{\rm g}$, then $\tau_{870 \mu\rm{m}} = \Sigma^{\rm up}_{\rm d} \chi_{870 \mu\rm{m}} > 1$, and $\Sigma_{\rm d} \gtrsim 6/\chi_{870 \mu\rm{m}}$, where $\chi_{870 \mu\rm{m}}$ is the average extinction coefficient at $\lambda = 870 \ \mu$m of the dust grains at $z > H_{\rm g}$.

The extinction coefficient at $\lambda = 870 \ \mu$m for grains with $a_{\rm max} =  100 \ \mu$m is $1.87$ cm$^{2}$ g$^{-1}$
(Figure 4 of \cite{Carrasco_2019}). Using this value as the average extinction coefficient in the disk surface, the dust column density needed for the disk to become optically thick with these grains is $\Sigma_{\rm d} \gtrsim 3.21$ g cm$^{-2}$. Typically, this dust surface density can only be reached in the most inner region of protoplanetary disks. 
For example, since the surface density of the gas is assumed to be 100 times larger than that of the dust, the above condition can be written in terms of the gas surface density as $\Sigma_{\rm g} \gtrsim  321$ g cm$^{-2}$, however, the gas surface density of a disk with a mass of 0.3 $M_{\odot}$ (the upper disk mass limit of a solar mass star \citep{Shu_1990}) and radius of 100 au is $\Sigma_{\rm g} =  4259 (\varpi / \rm au)^{-1}$ g cm$^{-2}$. Thus, only the most inner region with radii $\varpi< 13.3$ au has the required surface density such that the disk surface can become optically thick with grains of 100 $\mu$m in its surface.

Note that this radius is only a lower limit since the local dust-to-gas mass ratio could be enhanced by dust trapping and/or radial migration. However, independently of the assumed disk model, the small mass fraction available at the disk surface and the small value of the extinction coefficient for grains of 100 $\mu$m (compared with mm or cm grains), makes it difficult for settling to explain the disagreement between the inferred grain sizes from the polarization and the spectral index, specially at large disk radii.

In the case where the dust surface density is large enough to satisfy the above condition ($\Sigma_{\rm d} \gtrsim 3.21$ g cm$^{-2}$), the largest grains can be hidden by the optically thick disk surface and one would not recover the spectral index corresponding to the largest particles in the disk mid plane. For example, note from Figure (\ref{fig:alpha_settling}) that even when the global maximum grain size is 1 cm in all the models, these grains concentrate around the mid plane, with a scale height that depends on the $\alpha_t$ parameter. Then, the spectral indices of very settled models (small $\alpha_t$) do not match the spectral index for grains of 1 cm. The spectral index of 1 cm grains is recovered only for large values of $\alpha_t$, when all the dust grains are well-mixed.

\begin{figure}
\centering
\includegraphics[scale=0.7]{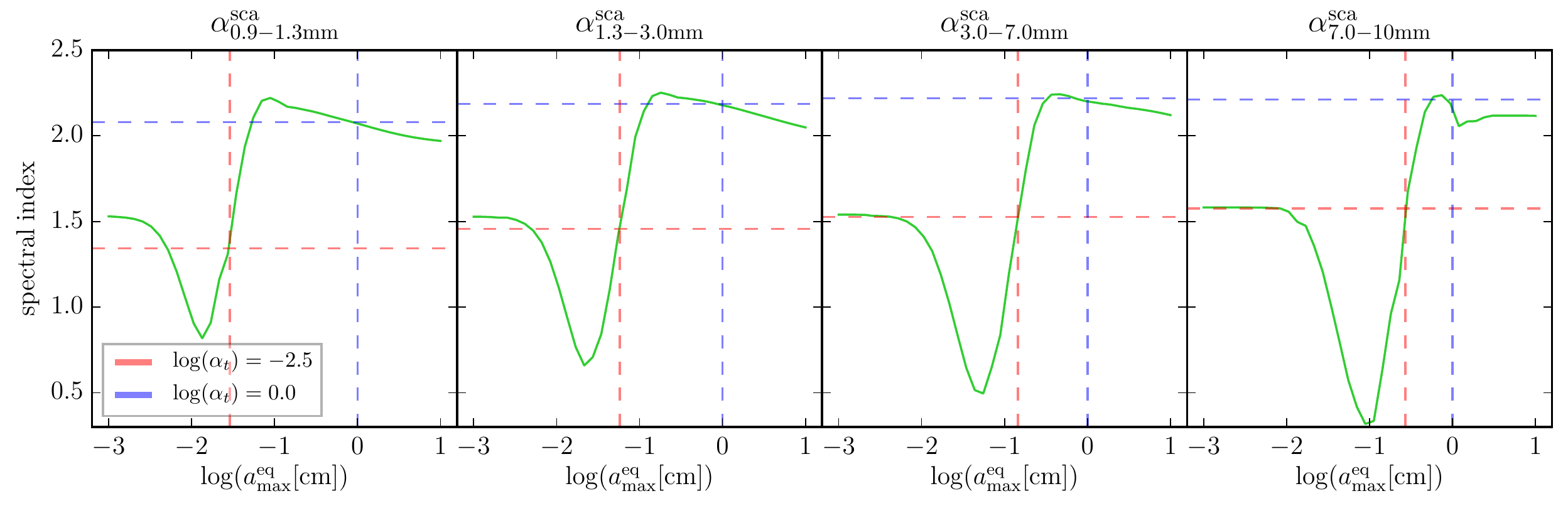}
\caption{Equivalent maximum grain size of a well-mixed disk model (green lines) that has the same spectral index of a settled disk with $\alpha_t = -2.5$ (horizontal red lines) and $\alpha_t = 0$ (horizontal blue lines).}
\label{fig:amax_settling}
\end{figure}

One could ask which is the equivalent well-mixed disk that has the same spectral index than a settled optically thick disk?
Green lines of Figure (\ref{fig:amax_settling}) show the spectral indices in the optically thick regime as a function of the maximum grain for disks with constant maximum grain size above the mid plane (i.e. no-settled disks). The spectral index is indicated in the top of each panel and the wavelength increases from left to right. These curves are the same than those shown in middle panels of Figure (\ref{fig:Temperature}) in the optically thick regime and with a temperature gradient. Also, we computed the spectral indices that correspond to settled disks with maximum grain size of 1 cm in the mid plane but with different settling degree $\log (\alpha_t) = -2.5, 0.0$, which are shown as red and blue horizontal lines, respectively. The vertical lines of the same color in each case is the intersection between the horizontal lines and the green curve. The intersection represents the equivalent well-mixed disk model that has the same spectral index than the settled model. 

One can see  that, for a high value of the turbulent parameter $\log (\alpha_t) = 0$, which corresponds to a well-mixed model, one recovers 
 $a_{\rm max} = 1$ cm, as expected. However, for highly settled disks with 
  $\log (\alpha_t) = -2.5$, the equivalent well-mixed model has a smaller grain size, due to the opacity of the upper layers of the disk that hide the large grains in the mid plane. Thus, in the latter models, based on the spectral indices, one would underestimate the maximum grain size. Table (\ref{FIG:amax_settling}) summarizes the equivalent well-mixed model at the different spectral indices. From this table, we note that the larger the wavelengths where the spectral index is measured, the larger the inferred maximum grain size. Furthermore, there is a heuristic factor between the equivalent maximum grain size and the shortest wavelength of each spectral index given by $a_{\rm max}^{\rm eq} \sim \lambda/\pi$.

\begin{table}
\centering
\begin{tabular}{cc}
\hline
spectral index & $a_{\rm max}^{\rm eq}$ \\
\hline
$\alpha_{0.9-1.3 \mathrm{mm}}^{\mathrm{sca}}$ & 290 $\mu$m \\
$\alpha_{1.3-3.0 \mathrm{mm}}^{\mathrm{sca}}$ & 580 $\mu$m \\
$\alpha_{3.0-7.0 \mathrm{mm}}^{\mathrm{sca}}$ & 1.5 mm \\
$\alpha_{7.0-10\mathrm{mm}}^{\mathrm{sca}}$   & 2.7 mm \\
\hline
\end{tabular}
\caption{Equivalent maximum grain size of a well-mixed disk model that has the same spectral index of a settled disk with maximum grain size of 1 cm in the mid plane.}
\label{FIG:amax_settling}
\end{table}

\newpage
\section{Spectral energy distribution}
\label{SEC:SED}

The inclusion of scattering in the radiative transfer equation modifies the spectral indices, as shown in the above sections. This means that the shape of the spectral energy distribution (SED) is modified with respect to the pure absorption case. In this section we explore the effects of the scattering on the SED for a given disk model with different inclination ($\mu = \cos \theta$) with respect to the plane of the sky.

The disk model corresponds to a gas surface density that decreases as $\Sigma_{\rm g} \propto \varpi^{-3/2}$, and has a magnitude of 1700 g cm$^{-2}$ at 1 au (the same than the minimum mass solar nebulae \citep{Weidenschilling_1977}) and a constant dust-to-gas mass ratio of 1/100 . The disk is assumed to be in hydrostatic equilibrium and the maximum grain size is set to $a_{\rm max} = 1$ mm in all the disk. We obtain the SED by solving numerically the radiative transfer equations along different lines of sight.

\begin{figure}
\centering
\includegraphics[scale=0.9]{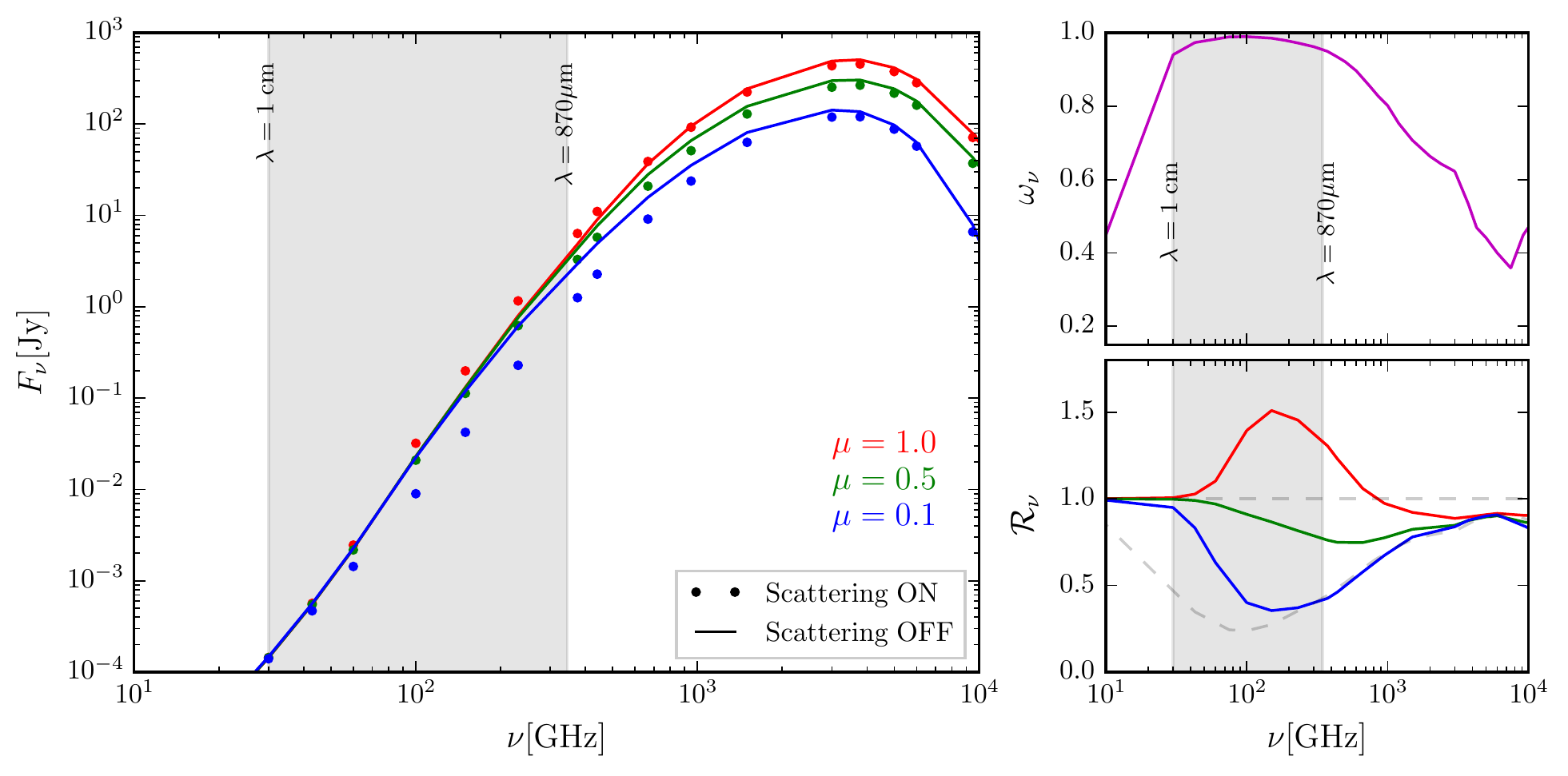}
\caption{Left panel: SED for a protoplanetary disk model (see text) viewed from different line of sights, from almost face-on (red) to edge-on (blue). The dot and solid lines are the models where the scattering is turn on and turn off respectively. Top right panel: Albedo as a function of the frequency for a maximum grain size of 1 mm. Bottom right panel: Intensity ratios $\cal{R}_{\nu}$ between the scattering and non-scattering fluxes for the different inclination angles. The maximum grain size of this model is set to $a_{\rm max} = 1$ mm in all the disk.}
\label{FIG:SED}
\end{figure}

The left panel of Figure \ref{FIG:SED} shows the SED for different inclination angles, from face-on (red) to edge-on (blue). In all the cases, the points represent the models that include scattering in the radiative transport. The solid lines are the models where only the true absorption is included. The albedo properties are shown in the top right panel. The bottom right panel show the ratio $R_{\nu}$ for the different inclination angles. The two reference dashed lines are: ${\cal R}_{\nu} =1$, and the optically thick limit (equation \ref{EQ:Optical_thick_limit}). A gray area between $\lambda =  870 \ \mu$m and 1 cm is included for reference in all panels, where many protoplanetary disks have been observed in the last years with ALMA and VLA.

In the mm range, the ratio ${\cal R}_{\nu}$ is larger than 1 (due to $\log (\tau_{\kappa_{\nu}})$ is between -2 and 0 in all the disk) for the face-on disk. However, ${\cal R}_{\nu}$ decreases for higher inclination angles due to the optical depth increases. These are same properties shown in Figure (\ref{FIG:Ratio_Intensities}), where for a constant and high albedo $\omega_{\nu}$, the ratio ${\cal R}_{\nu}$ decreases from left to right as the the optical depth increases. 
Out of the mm range, the ratio ${\cal R}_{\nu} \sim 1$ because the albedo is not as large as in the mm range. 
The largest decrease of the SED when scattering is included occurs for the largest inclination angle  $\mu=0.1$ (blue solid line in the bottom right panel). The deficit of emission in the edge-on disk (blue curves) is compensated by the increase of the emission in the face-on disk (red curves); i.e. scattered light preferentially escapes from the optically thinnest direction.

Note that, since 1 mm fluffly dust grains preferentially scatter the photons in the forward direction (i.e. the effective albedo tends to 0), the scattering effects can be neglected for these grains \citep{Tazaki_2019}. Then, the SEDs in Figure (\ref{FIG:SED}) can also be interpreted as that corresponding to 1 mm compact dust grains (scattering-on) and 1 mm fluffy dust grains (scattering-off).

\section{Apparent excess emission at 7 mm}
\label{SEC:7mm}
Excess emission at 7 mm has been reported in several disks around T Tauri stars (e.g. the disks around Di Cha, T Cha, Sz 32 \citep{Ubach_2017}) and Herbig AeBe stars (e.g, the disks around HD 35187, HD 142666, HD 169142 \citep{Sandell_2011}). In these disks, the observed intensity at 7 mm is larger than the intensity expected from an SED extrapolation using the ALMA wavelengths. The excess emission, which in some cases is a factor of 2 or larger, cannot be explained by the flux calibrator uncertainty of the VLA at 7 mm, where the  flux uncertainty  is $\sim 10$\% for a single epoch observation.

Many authors have interpreted the excess emission as optically thick free-free emission from a compact ionized gas (e.g. \citealt{Macias_2017}), free-free emission from ionized winds \citep{Sandell_2011}, or emission from spinning dust \citep{Hoang_2018}.  The latter occurs due to the fast rotation of polycyclic aromatic hydrocarbons (PAHs) or silicate grains with sizes of some nanometers. The flux from the spinning dust emission at low frequencies ($\nu < 60$ GHz) can be one order of magnitude larger than that of the dust termal emission for Herbig AeBe and T Tauri stars depending on the dust size distribution of these nano particles. Although this effect has not been confirmed, the existence of the nano silicate grains should also match with the presence of silicate features at smaller wavelengths ($\lambda \sim 10\ \mu$m). It is not clear to the date what is the main physical mechanism that can produce the 7 mm excess.

Here, it is shown that the effect of scattering in the isothermal case, can produce an {\it apparent excess emission} at $\lambda = 7$ mm when interpreted as a pure absorption case. Consider a disk with mm or cm size grains, where the albedo is large at mm wavelengths ($\omega_{\nu}  \gtrsim 0.6$). Consider further a typical case, where  the disk is optically thick at ALMA wavelengths but optically thin at 7 mm observed with the VLA. 
In these conditions, the emergent intensity decreases at the ALMA wavelengths and increases at 7 mm compared with the pure absorption case. This combined effect could be interpreted as an excess emission at 7 mm if the ALMA emission is extrapolated to 7 mm.

The effect of the maximum grain size on the apparent excess at 7 mm can be studied based on the spectral indices in Figure (\ref{FIG:spectral_index}) as follows: In the typical pure absorption case, the spectral indices at sub-mm and small mm wavelengths varies from $\alpha^{\rm abs}_{\lambda_1,\lambda_2} = 2$ (at small optically thick wavelengths) to $\alpha^{\rm abs}_{\lambda_1,\lambda_2} = 2+\beta_{\rm \kappa_\nu}$ (at large optically thin wavelengths); then, as $\beta_{\kappa_{\nu}} > 0$, the ratio $\alpha^{\rm abs}_{1.3-3.0 \rm mm}/\alpha^{\rm abs}_{3.0-7.0 \rm mm}$ is always less than 1. This does not always occur in the scattering case, where the ratio $\alpha^{\rm sca}_{1.3-3.0 \rm mm}/\alpha^{\rm sca}_{3.0-7.0 \rm mm}$ could be larger than 1 depending on the optical depth regime and the maximum grain size.

\begin{figure}
\centering
\includegraphics[scale=0.9]{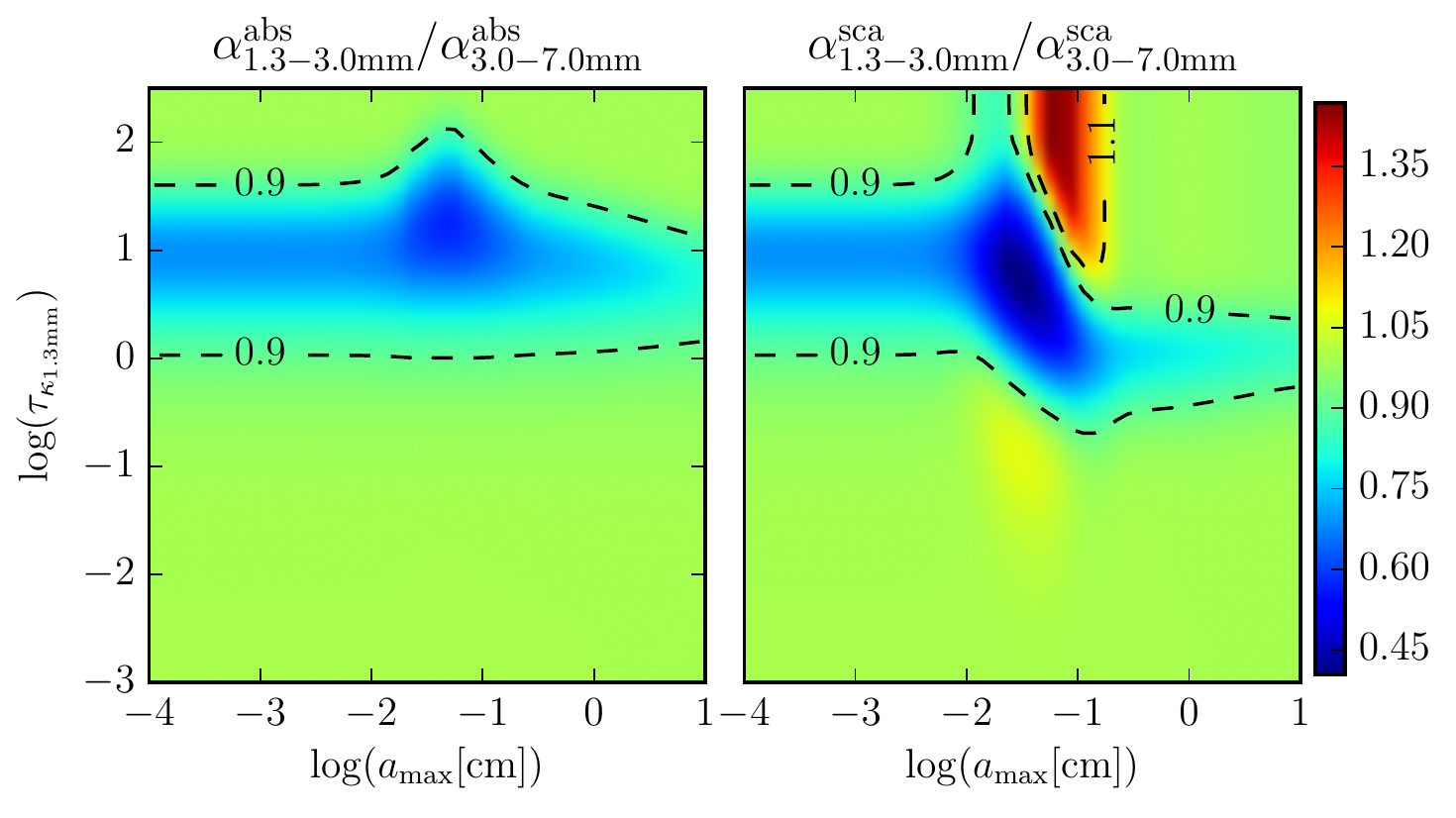}
\caption{Ratio between the spectral indices in the scattering and absorption cases (see Figure \ref{FIG:spectral_index}) at mm wavelengths as a function of the optical depth at $\lambda = $1.3 mm and the maximum grain size. For reference, the dashed lines shows the region where the ratio is 0.9 and 1.1.}
\label{FIG:Comparacion}
\end{figure}

Figure (\ref{FIG:Comparacion}) shows the ratio between $\alpha_{1.3-3.0 \rm mm}$ and $\alpha_{3.0-7.0 \rm mm}$ in the true absorption case (left panel) and the scattering case (right panel) as function of the optical depth at $1.3$ mm and the maximum grain size. Isocontours where the ratio is 0.9 and 1.1 are shown as reference as dashed lines. In the absorption case, the ratio is 1 in the optically thin regime (where both spectral indices are given by $2+\beta_{\kappa}$) and in the optically thick regime (where both spectral indices are $2$). For intermediate optical depths, the spectral index at smaller wavelengths is smaller than at longer wavelengths, thus the ratio is always smaller than 1. 

In the scattering case, the ratio is also 1 in the optically thin regime. For very optically thick disks ($\log (\tau_{\kappa_{1.3 \rm mm}}) \gtrsim 1.5$), the ratio is also 1 for very small grains (where the albedo is small) and for very large grains (where the albedo is approximately constant ($\beta_{\omega_\nu} \sim 0$) and the scattering equally affects all the wavelengths (equation \ref{EQ:Optical_thick_limit}). However, the region between $-1.5 \lesssim \log (a_{\rm max}[\mathrm{cm}]) \lesssim -1$ and $ \log (\tau_{\kappa_{1.3 \rm mm}}) \gtrsim 1$ has a ratio larger than 1. The SED of a disk with the latter properties would look anomalous and the emergent intensity at 7 mm could be interpreted as an excess emission if the scattering effects are not taken into account.

Observational evidence where the ratio between spectral indices is larger than 1 has been found for example in the FU Ori Disk (see e.g., Table 2 and Section 4.4.1 of \cite{Liub2019}).

\begin{figure}
\centering
\includegraphics[scale=0.9]{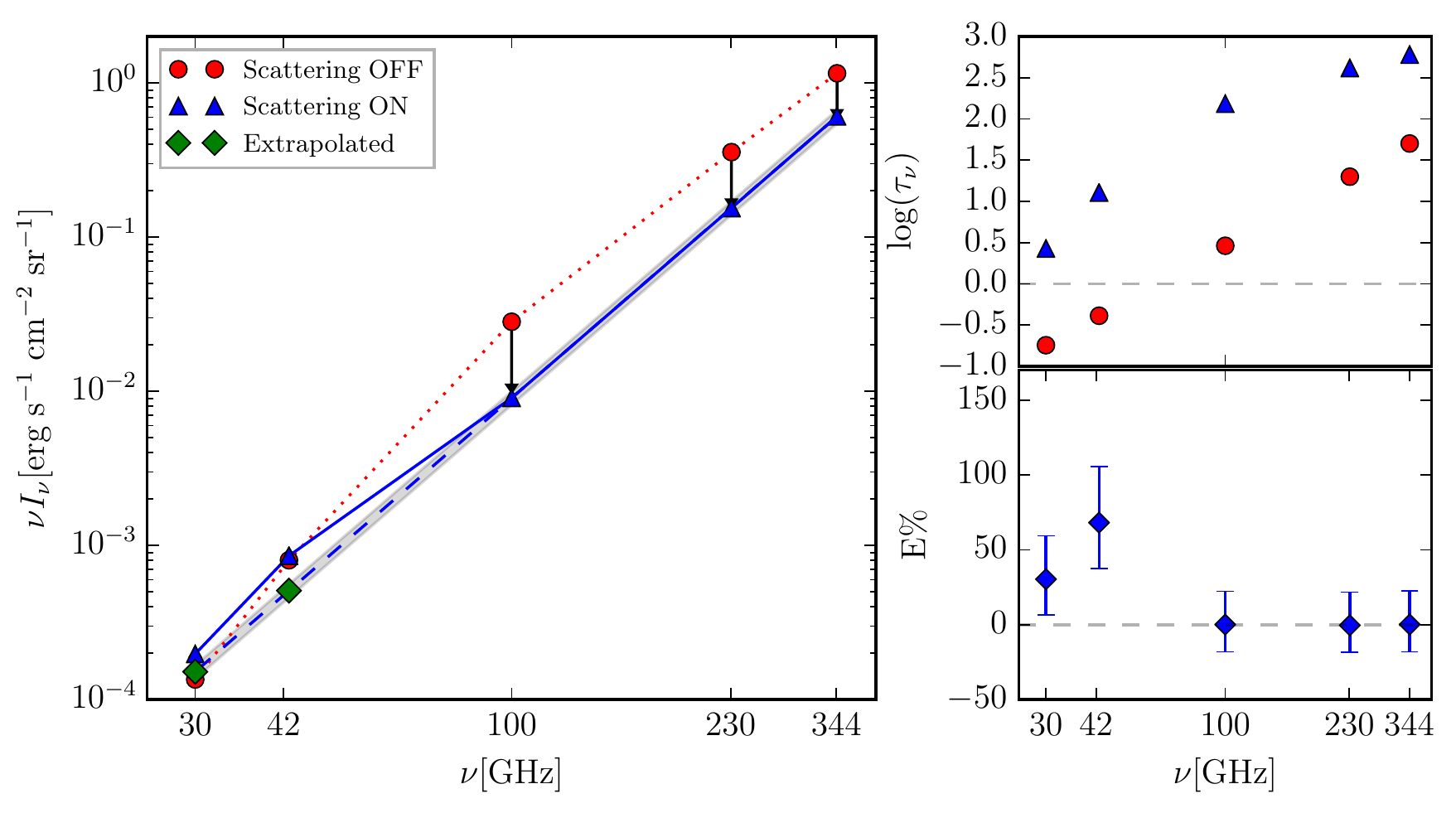}
\caption{Scattering effects on the SED for a disk with maximum grain size $a_{\rm max} = 1$ mm and $\log (\tau_{\kappa_{1 \mathrm{mm}}}) $ = 1.3. Left panel: SED at millimeter wavelengths for a pure absorption model (red dots) and taking into account the scattering (blue triangles). The blue crosses are the extrapolated VLA intensity using the ALMA frequencies. Top right panel: optical depths associated to the disk at different wavelengths in the scattering off (red dots) and scattering on (blue triangles) models. Bottom right panel: Excess emission at different wavelengths.}
\label{FIG:Excess}
\end{figure}

In Figure (\ref{FIG:Excess}) we  explore the effects of the scattering on the 7 mm emission interpretation for a disk with $a_{\rm max} = 1$ mm and $\log (\tau_{\kappa_{1.3 \rm mm}}) = 1.3$ using equations (\ref{EQ:Intensity_scattering}) and (\ref{EQ:Intensity_absorption}). The SED of this disk model is shown in the left panel. The red dots are the pure absorption emission case (scattering OFF), and the blue triangles are the scattering case (scattering ON). The black arrows show the difference between both cases. Note that the emission in all the ALMA wavelengths decreases, while at VLA wavelengths the emission slightly increases by a factor of $\sim 1.1$.

Then, if we wrongly interpret the millimeter emission as a pure absorption case (which is the usual assumption in many papers) and extrapolate the emission based on the ALMA frequencies ($\nu = 100, 230, 344$ GHz) to smaller frequencies ($\nu = 30, 42$ GHz), the observed intensities at these small frequencies would seem to have an excess emission compared with the extrapolated emission (dashed blue line and green diamonds) due to the combined effect of the decrease of the emission at optically thick wavelengths and the increase of the emission at optically thin wavelengths. 

The flux calibration error at ALMA and VLA is around $\sim 10 \%$. The width of the extrapolated line takes into account the propagation due to this uncertainty. Even when this error is taken into account, it is not enough to explain the excess by calibration errors.

Top right panel of this Figure shows the optical depth associated to the scattering (blue triangles) and pure absorption cases (red dots). The bottom right panel shows the emission excess defined as 
\begin{equation}
\rm E \% = \left(\frac{I_{\nu}^{\rm sca}  - I_{\nu}^{\rm ext}}{I_{\nu}^{\rm ext}} \right) \times 100,
\label{EQ:Excess}
\end{equation}
where $I_{\nu}^{\rm ext}$ is the extrapolated intensity (green diamonds).
The vertical error bars correspond to the propagation of the uncertainties in the flux calibration in the ALMA and VLA observations. Note that at 42 GHz, the emission excess is $\sim 60$\%. 

Finally, we note that,  even if the 7 mm emission slightly increases in the scattering case, the main reason for the apparent 7 mm excess is the decrease of the emission at optically thick millimeter (ALMA) wavelengths due to the scattering effects.  
This deficit of the emission at optically thick ALMA  wavelengths also lead to \cite{Zhu_2019} to propose that the disks are more massive than expected when the observations are wrongly interpreted as a pure absorption case.

A multi-wavelength modelling of sources with an inferred excess emission at 7 mm is necessary to determine if the latter can be explained by scattering. Nevertheless, this is beyond the scope of this paper.

\section{Conclusions}
\label{SEC:conclusions}

As pointed out by  \cite{Miyake_1993}, the
 scattering coefficient is much larger than the absorption coefficient at millimeter wavelengths if the dust grains are mm-cm sized, as expected in protoplanetary disks. In this case, the albedo is close to 1 for grains with a size $a \sim \lambda/2\pi$. Therefore, a realistic radiative transfer solution in protoplanetary disks should not neglect the scattering effects. 
 
 In this work, we have discussed the differences in the properties of the emergent emission of face-on protoplanetary disks when scattering is or is not included. In particular, we address the difference between the spectral indices, and the spectral energy distribution, as a function of the albedo  and the optical depth. Given the opacity properties computed with the Mie theory for compact and spherical grains, we discuss how the maximum grain size can be inferred in the scattering and true absorption cases. 
We explore the effect on the spectral indices when the disk is not vertically isothermal, but has a vertical structure given by the heating of the central star and viscosity. The effects of dust settling on the spectral indices is also explored. 
We find that in highly settled opaque disks, the maximum grain size inferred from the spectral indices is underestimated because large grains in the disk mid plane are hidden by the small grains in the disk surface.

Instead of looking at specific spectral indices, one can study the effect of scattering in the shape of the spectral energy distribution.
One can see that larger inclination angles of the disk with respect to the plane of the sky imply a larger decrease of the flux at millimeter wavelengths in the scattering case. This occurs because more inclined disks have a larger optical depth along the line of sight. The deficit of the edge-on emission is compensated by the increase of the face-on emission.

The modification of the spectral indices at optically thick wavelengths in the scattering case could provide  an alternative explanation to the 7 mm excess emission reported in some disks around T Tauri and Herbig AeBe stars.

Quantitatively, our main results are summarized as follows:
\begin{enumerate}
\item Scattering modifies the emergent intensity (compared with the true absorption case) in protoplanetary disk with large albedo ($\omega_{\nu} \gtrsim 0.6$). For a vertically isothermal slab in the optically thick regime, the emergent intensity can decrease by a factor of 4. For intermediate optical depths $-2 \lesssim \log (\tau_{\kappa_\nu})\lesssim -1 $, it increases by a factor of 2, while in the optically thin regime the emergent intensity is not modified from the true absorption case.

\item In the isothermal case, the changes in the emergent intensity due to the scattering effects modify the inferred spectral indices at millimeter wavelengths compared with the true absorption case. The spectral indices in the scattering and true absorption case coincide for all wavelengths only in optically thin case ($\log (\tau_{\kappa_{1.3 \rm mm}}) \lesssim -2.5$), but they do not coincide at large optical depths. In particular, when scattering is included, spectral indices smaller than 2 can be obtained for optically thick disks  and dust grains with sizes between $\sim 100\ \mu$m and $\sim 1$ mm. 

\item In addition to the scattering effects, the vertical temperature structure modifies the spectral indices. The temperature gradient decreases the spectral indices in the optically thick regime because larger wavelengths can penetrate deeper in the disk, where the temperature increases due to the viscous heating. When scattering is included, spectral indices close to 0 can be reached for very optically thick disks if the vertical structure of the temperature is taken into account and the grain size is $\sim 1$ mm.

\item Settling also modifies the spectral indices in the optically thick regime because small grains in the upper disk layers can hide the large grains around the mid plane. The modification depends on the degree of settling, determined by the magnitude of the turbulent parameter  $\alpha_t$.
If the dust surface density is larger than $\Sigma_{\rm d} \gtrsim 6/\chi_{\nu}$, the emission of the large grains in the disk mid plane can be hidden by the small dust grains in the upper layers of the disk. In particular, at $\lambda = 870 \ \mu$m, the dust surface density needs to be larger than $\gtrsim$ 3.21 g cm$^{-2}$. Without phenomena like dust radial migration and/or dust 
trapping, this condition can only be satisfied in the inner regions of the disks ($\varpi \lesssim 14$ au). Thus, the large dust column density needed to explain by settling the disagreement between the inferred grains from the polarization method and the spectral index, 
is not expected to occur at large radii, unless dust is trapped in pressure maxima.
In the regions where the dust column density is large enough such that the large grains in the mid plane can be hidden, the inferred maximum grain size of very settled disks is smaller than the true maximum grain size. 
If the spectral index is computed at larger, optically thinner wavelengths, the maximum grain size is better estimated.

\item Scattering modifies the shape of the spectral energy distribution (SED) when the albedo is large ($\omega_{\nu} \gtrsim 0.6$) and the optical depth is $\log (\tau_{\kappa_\nu} )  \gtrsim  -2$. The effects of the scattering on the SED depend on the inclination of the disk with respect to the plane of the sky because more inclined disks are more optically thick than face-on disks. 
In particular, the neglect of the scattering effects on the radiative transfer of protoplanetary disks can lead to a wrong interpretation of an apparent excess emission at optically thin millimeter wavelengths (e.g., $\lambda = 7$ mm). For example, the incorrect interpretation of the observed SED as a true absorption case, would imply an excess emission of $\sim 60$\% for a disk with $a_{\rm max} = 1$ mm and $\log(\tau_{\kappa_{1.3 \rm mm}}) = 1.3$. 
Thus, the 7 mm excess reported in several sources, could be explained by optically thick disks ($\log(\tau_{\kappa_{1.3 \rm mm}}) \gtrsim 1$) and dust grain sizes between 300 $\mu$m $\lesssim a_{\rm max} \lesssim 1$ mm. This possibility needs to be explored by multi-wavelength modelling of observed sources where this excess has been reported.
\end{enumerate}

\textit{Acknowledgements}\\
A.S. and S.L. acknowledge support from PAPIIT-UNAM IN101418 and CONACyT 23863. We thank useful comments from an anonymous referee that helped clarified some aspects of the paper.

\appendix
\section{Spectral Indices for cold disks}
\label{App:RJeans}
The spectral indices for the isothermal slab in Section (\ref{SEC:spectral_index}) were computed for a temperature $T=100$ K, where the emission at millimeter wavelengths is in the Rayleigh-Jeans regime.
However, for colder disks, the spectral indices are expected to change because the Rayleigh-Jeans approximation is no longer valid.

\begin{figure}
\centering
\includegraphics[scale=0.65]{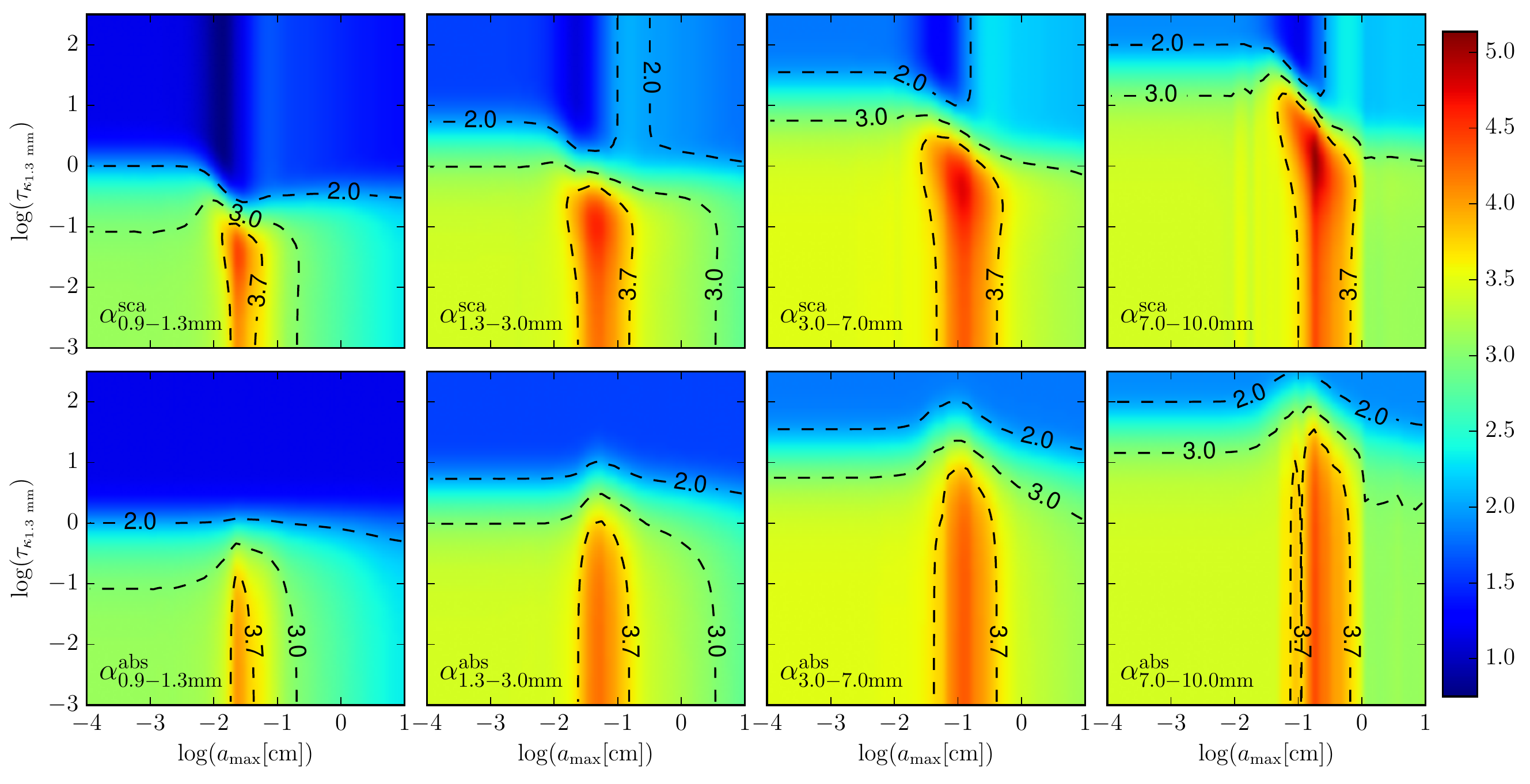}
\caption{Spectral indicies in the mm range as a function of the optical depth a 1.3 mm and the maximum grain size. The slope of the particle size distribution is fixed to $p=3.5$ and the temperature is $T=10$ K. In the top panels the scattering effects are taken into account, while in the bottom panels they are ignored. The color bar is the same in all panels.}
\label{FIG:Spectral_index_cold}
\end{figure}

Figure (\ref{FIG:Spectral_index_cold}) shows 
the same spectral indices as Figure (\ref{FIG:spectral_index}), but at a temperature of $T=10$ K. Note that the spectral indices between 3.0 - 7.0 mm and 7.0-10.0 mm do not have strong changes compared with the disk at $T=100$ K because the Rayleigh-Jeans regime is still valid at
 these long wavelengths. However, the spectral indices between 0.9-1.3 mm and 1.3-3.0 mm of Figure (\ref{FIG:Spectral_index_cold}) are, in
 general, lower than those shown in Figure (\ref{FIG:spectral_index}). In particular, the spectral index $\alpha^{\rm sca}_{0.9-1.3 \rm}$ can reach values below of 1 for grains with $a_{\rm max} \sim 100 \ \mu$m and optical depths $\log (\tau_{\kappa_{1.3 \rm mm}}) >$ 0. 
This occurs because the peak of the black body radiation at $T = 10$ K is around $\lambda_{\rm peak} \sim 300 \ \mu$m, thus, 
it strongly affects the slope of the SED at $\lambda = 870 \ \mu$m.

\end{document}